\documentclass[prb,aps,twocolumn,superscriptaddress,amsmath,amssymb,floatfix]{revtex4}
\usepackage{graphics}
\usepackage{psfrag}

\usepackage{epsf,graphicx,xspace,epsfig,pstricks}

 \usepackage{color}

\usepackage{amssymb,graphicx,xspace,color}  
\usepackage{epsf,epsfig,psfrag}
\usepackage{amsmath}
\usepackage{url}
\begin{document}

\title{
  Generalized diagonalization scheme for
  many-particle systems
}

\author{Steffen Sykora}

\affiliation{
Institute for Theoretical Physics and W\"{u}rzburg-Dresden Cluster of Excellence ct.qmat,
  Technische Universit\"{a}t Dresden, 01069 Dresden, Germany
}

\affiliation{Leibniz-Institute for Solid State and Materials Research, IFW Dresden, 01069 Dresden, Germany}

\author{Arnd H\"ubsch}

\affiliation{
Institute for Theoretical Physics and W\"{u}rzburg-Dresden Cluster of Excellence ct.qmat,
  Technische Universit\"{a}t Dresden, 01069 Dresden, Germany
  }

\author{Klaus W.~Becker}

\affiliation{
Institute for Theoretical Physics and W\"{u}rzburg-Dresden Cluster of Excellence ct.qmat,
  Technische Universit\"{a}t Dresden, 01069 Dresden, Germany
}

\date{\today}

\begin{abstract}
  Despite the advances in the development of numerical methods 
  analytical approaches still play the key role on the way towards a deeper 
  understanding of many-particle  systems. In this regards, 
  diagonalization schemes for Hamiltonians represent an important 
  direction in the field. Among these techniques the 
 method, presented here, might be that 
  approach with the widest range of possible applications: We demonstrate that both stepwise and continuous unitary transformations to diagonalize the many-particle Hamiltonian
  as well as perturbation theory and also non-perturbative treatments can be 
  understood within the same theoretical framework.
   The new method is based on the introduction of generalized projection operators 
  and allows to develop a renormalization scheme which is used to evaluate directly the  physical quantities 
  of a many-particle system. The applicability of this approach is shown for two important elementary  many-particle 
  problems.
\end{abstract}

\maketitle

\section{Introduction}
\label{I}
During the last three decades the investigation of phenomena related to
strongly interacting electrons has developed to a central field of condensed
matter physics. In this context, high-temperature superconductivity and
heavy-fermion behavior are maybe the most important examples. It has been
clearly turned out that such systems require true many-body approaches that
properly take into account the dominant  electronic correlations. 

In the past, many powerful numerical methods like exact diagonalization
\cite{ED}, numerical renormalization group \cite{NRG,NRG_2}, Quantum Monte-Carlo
\cite{MC}, the density-matrix renormalization group (DMRG) \cite{DMRG,DMRG_2}, or the
dynamical mean-field theory (DMFT) \cite{DMFT,DMFT_2} have been developed to study strongly
correlated electronic systems. In contrast, only very few analytical
approaches are available to tackle such systems. In this regard,
renormalization schemes for Hamiltonians developed in the nineties of the last
century \cite{GW_1993, GW_1994, W_1994, Kehrein_2006} represent an important new direction
in the field where renormalization schemes are implemented in the Liouville
space (that is built up by all operators of the Hilbert space). Thus, these
approaches can be considered as further developments of common renormalization
group theory \cite{RG} which is based on a renormalization within the Hilbert 
space. 

In the present study we discuss a generalized diagonalization method that shares some basic concepts with the renormalization schemes for Hamiltonians mentioned above 
\cite{GW_1993, GW_1994, W_1994, Kehrein_2006}. All these approaches including the present one  
generate diagonal Hamiltonians by applying a sequence of unitary
transformations to the initial Hamiltonian of the physical system. However,
there is one distinct difference between these methods: Both similarity
renormalization \cite{GW_1993,GW_1994} and Wegner's flow equation method
\cite{W_1994,Kehrein_2006} start from a continuous formulation of the unitary transformation
by means of a differential form. In contrast, our method is based on discrete
transformations so that a direct link to perturbation theory can be provided. 

This paper is organized as follows: 
In the next section (Sec.~\ref{II}) we discuss the basic concepts of our method. We introduce 
projection operators in the Liouville space that allow the definition 
of an effective Hamiltonian. If these ingredients are combined with unitary 
transformations one obtains a new renormalization scheme which is based on a stepwise elimination of interactions. We derive the corresponding formalism in great detail in Sec.~\ref{II.A} and illustrate the introduced steps  for the case of an exactly solvable model in Sec.~\ref{App A1}.
In this context the relation to Wegner's flow equation method \cite{W_1994,Kehrein_2006} is also shown. It turns out that the latter method can be understood within the more general framework of our approach
by choosing a complementary unitary transformation to generate the effective 
Hamiltonian. For demonstration, the exactly solvable  model is treated  with this 
approach, too. 

In Sec.~\ref{III} we explain the technique to analyze many-particle systems with interactions. This is presented  for two rather elementary examples:  the Holstein model and the extended Falicov-Kimball model. Both models are prototypes   of  systems where the renormalization of all parameters can be simultaneously taken into account. We give a detailed description how the renormalization equations are derived and numerically evaluated. 
Furthermore, the method to calculate expectation values within our approach is discussed in detail. We present the corresponding numerical results and discuss different parameter regimes of the one-dimensional Holstein model in the metallic state. 
It is well-known that the Holstein
system undergoes a quantum phase transition from a metallic 
to a Peierls distorted state if the electron-phonon coupling exceeds a 
critical value. However, first we discuss the 
crossover behavior between the adiabatic and 
anti-adiabatic case for the metallic state. All physical properties are shown to strongly depend on the ratio of  
initial parameters of  the system. 
It is also demonstrated in Sec.~\ref{III}   that our method can  be used to study models which include fermion-fermion interaction from the beginning and not necessarily provided by other degrees of freedom. This is shown in the example of the extended Falicov-Kimball model. The corresponding Hamiltonian is considered to capture the anticipated 'BCS-Bose-Einstein condensate crossover' scenario. We show how one-particle spectral functions can be evaluated and to what extent the results can be used to understand the behavior of spectral weight transfer which appears under a variation of the Coulomb interaction strength.

In Sec.~\ref{X} a short summary of the basic concepts of our technique is presented and the advantages over other many-particle approaches is discussed.

\section{Generalized diagonalization scheme}
\label{II}

Let us begin with the basic concepts of our method, which were partially introduced in Ref.~\cite{BHS_2002}. Based on the introduction of generalized projection 
operators our aim is to derive effective Hamiltonians which are diagonal. However, instead of \textit{states} as in usual renormalization group approaches
 {\it transition operators}  are integrated out. In this way,  a renormalization scheme is established,
which allows to diagonalize  many-particle 
Hamiltonians.

\subsection{Basic concepts}
\label{II.A}

The method
starts from the decomposition of a given many-particle Hamiltonian into
an unperturbed part $\mathcal H_0$ and into a perturbation $\mathcal H_1$, 
\begin{eqnarray}
\label{2.1}
\mathcal{H} = \mathcal{H}_{0} + \mathcal{H}_{1} .  
\end{eqnarray}
Without loss of generality let us assume that no part of $\mathcal{H}_{1}$ 
commutes with $\mathcal{H}_{0}$. Thus, the perturbation 
$\mathcal{H}_{1}$  accounts for transitions between eigenstates of
$\mathcal{H}_{0}$ with {\it non-zero} transition
energies only.  Usually, the presence of 
$\mathcal{H}_{1}$ prevents an exact solution
of the eigenvalue problem of $\mathcal{H}$. In spite of this property, the method allows 
to construct from $\mathcal H$ an effective Hamiltonian $\tilde{\mathcal H}$, which is solvable and  
allows to evaluate all relevant physical quantities. 

\begin{figure}[h]
  \centering
\includegraphics[width=0.9\columnwidth]{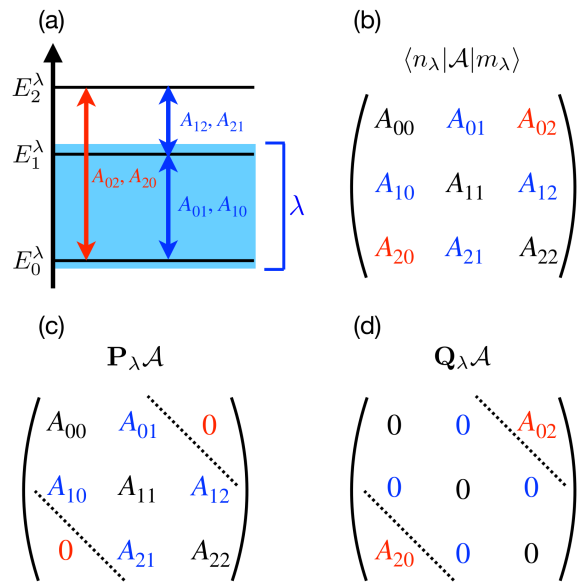}
  \caption{
    (Color online) Definition  of the projectors $\mathbf P_\lambda$ and $\mathbf Q_\lambda$ for a simple toy model consisting of three different eigenstates $| n_{\lambda} \rangle$ of ${\cal H}_{0,\lambda}$. (a) Corresponding eigenvalues $E_n^\lambda$  and transition matrix elements of an observable $\mathcal A$. The energy range of the  cutoff $\lambda$ is indicated by the blue shaded region. 
 Zero-energy transitions of $\mathcal A$ between same eigenstates are not shown.    (b) Arrangement of the  matrix elements $A_{nm} =
  \langle n_{\lambda} | \mathcal{A} | m_{\lambda}\rangle$ shown in the  respective colors distinguishing the  different transition energies shown in panel (a). (c) Action of the  projector $\mathbf P_\lambda$. It cuts out the matrix elements $A_{02}$ and $A_{20}$ with the energy difference larger than $\lambda$. All other matrix elements with energy differences smaller than $\lambda$ are kept unchanged. (d) The orthogonal projector $\mathbf Q_\lambda$ cuts out the low-energy transition matrix elements. 
  }
  \label{Fig_projectors}
\end{figure}

An essential element of our diagonalization scheme is the choice of an appropriate
Hamiltonian $\mathcal{H}_{\lambda}$, which depends on a given energy cutoff $\lambda$. 
Thereby, $\mathcal H_\lambda$ should have the following properties:
\begin{enumerate}
\item[(i)]
Just as for $\mathcal H$, also $\mathcal H_\lambda$ can be decomposed into  
an unperturbed part ${\cal H}_{0,\lambda}$ and into a perturbative part 
${\cal H}_{1,\lambda}$,
\begin{eqnarray}
  \label{2.2}
  \mathcal{H}_{\lambda} = 
 \mathcal{H}_{0,\lambda} +   \mathcal{H}_{1,\lambda} ,
\end{eqnarray}
where both parts depend on $\lambda$.
 \item[(ii)] 
  The eigenvalue problem of $\mathcal{H}_{0,\lambda}$ is solvable,
  \begin{eqnarray}
  \label{2.3}
    \mathcal{H}_{0,\lambda} | n_{\lambda} \rangle = 
    E_n^\lambda | n_\lambda \rangle  ,
  \end{eqnarray}
  where $E_n^ \lambda$ and 
 $|n_{\lambda}\rangle$ are the eigenvalues and  eigenvectors of $\mathcal H_{0,\lambda}$. 
 As before, the perturbation $\mathcal H_{1,\lambda}$ accounts for transitions 
  between the eigenstates of $\mathcal H_{0,\lambda}$.  Fig.~\ref{Fig_projectors} illustrates this situation for an example system of three different eigenstates.
  \item[(iii)]
  $\mathcal{H}_{\lambda}$ is  
  constructed such that only transitions 
  with energies smaller than $\lambda$ are left (blue arrows in Fig.~\ref{Fig_projectors}(a)).  That is, all transitions
 with excitation energies larger  than $\lambda$ (red arrows) have already been 
 eliminated from $\mathcal H_{1,\lambda}$. 
  %
  \item[(iv)]
  $\mathcal{H}_{\lambda}$ should have the same eigenvalues as 
  the original Hamiltonian ${\cal H}$.
\end{enumerate}

As it turns out, the solvable eigenvalue problem of ${\cal H}_{0,\lambda}$ is crucial for the 
construction of Hamiltonian ${\cal H}_\lambda$, because it
is used to define generalized projection operators $\mathbf P_\lambda$ and $\mathbf Q_\lambda$, 
\begin{align}
  \label{2.4}
  \mathbf {P}_{\lambda} {\mathcal{A}} &=
  \sum_{m,n} | n_{\lambda} \rangle \langle m_{\lambda} | \, 
 A_{nm}^\lambda \Theta(\lambda -|E_n^\lambda -E_m^\lambda |), \\
  \mathbf{Q}_{\lambda} \mathcal A &= (\mathbf{1} - \mathbf{P}_{\lambda}) \mathcal A, \nonumber
\end{align}
where $A_{nm}^\lambda =
  \langle n_{\lambda} | \mathcal{A} | m_{\lambda}\rangle$ is the matrix of any operator $\mathcal{A}$ in the basis of the eigenstates of ${\cal H}_{0,\lambda}$. The action of these projectors is illustrated in Fig.~\ref{Fig_projectors}.
Note that  $\mathbf P_\lambda$ and $\mathbf Q_\lambda$ do not act on states as usual 
projectors in the Hilbert space. 
Instead, they act on operators $\mathcal A$ of the unitary space,  
and are examples for so-called super\-operators.
Due to the $\Theta$-function in Eq.~\eqref{2.4} $\mathbf{P}_{\lambda}$ projects on
that part of ${\mathcal{A}}$ which is composed of all transition operators 
$| n_{\lambda}\rangle \langle m_{\lambda} |$ with transition energies  
$| E_n^\lambda - E_m^\lambda| $ less than $\lambda$. By contrast, 
$\mathbf{Q}_{\lambda}$ projects onto the orthogonal part of
${\mathcal{A}}$ with transition energies larger than $\lambda$.
Note that the eigenstates $|n_{\lambda}\rangle$ and $|m_{\lambda}\rangle$ of $\mathcal{H}_{0,\lambda}$ 
not necessarily belong to low energies. Only their difference $| E_n^\lambda - E_m^\lambda|$ has to 
be smaller than $\lambda$.

Obviously property (iii) is fulfilled when ${\cal H}_\lambda$ obeys
\begin{eqnarray}
  \label{2.6}
  \mathcal{H}_{\lambda} = \mathbf{P}_{\lambda} 
  \mathcal{H}_{\lambda}
\quad \mbox{or} \quad
  \mathbf{Q}_{\lambda} \mathcal{H}_{\lambda} = 0,
 \end{eqnarray}
whereas property (iv) is realized when $\mathcal H_\lambda$ and 
${\cal H}$  are related by a unitary transformation, 
\begin{eqnarray}
  \label{2.7}
  \mathcal{H}_{\lambda} = 
  e^{X_{\lambda}}\; \mathcal{H}\;  e^{-X_{\lambda}} . 
\end{eqnarray}
Here $X_\lambda= -X_\lambda^\dag$ is the generator of the unitary transformation.  
Relation  \eqref{2.6} will be used below to fix the generator of the unitary transformation.

\begin{figure*}[t]
  \centering
\includegraphics[width=\textwidth]{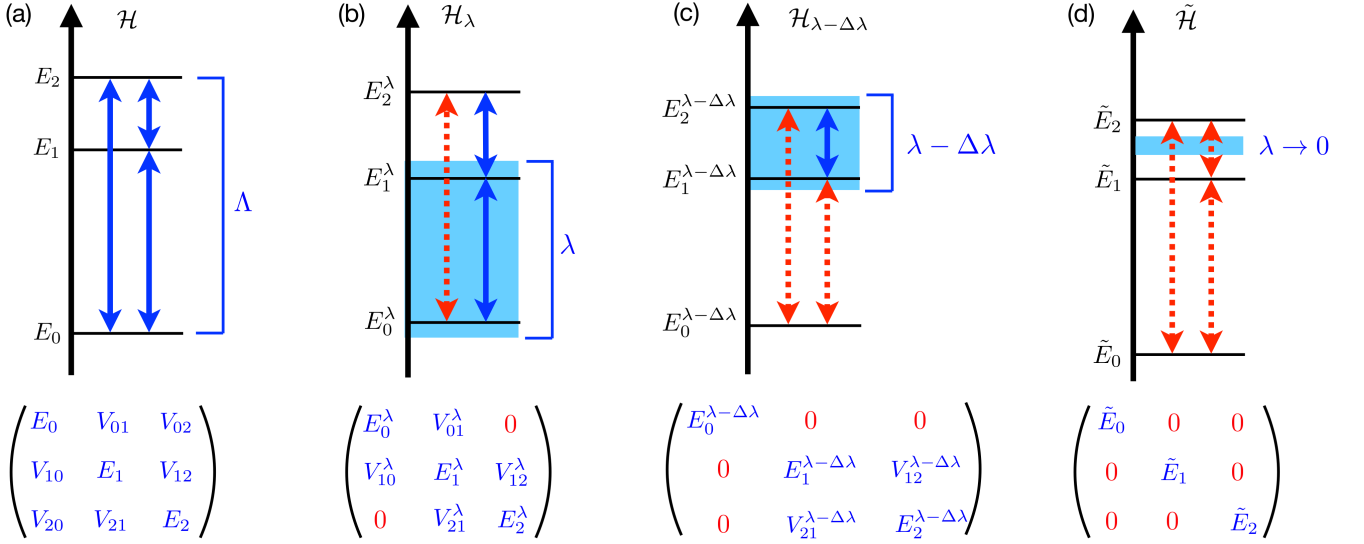}
  \caption{
    (Color online) Illustration of the diagonalization scheme for the example system of Fig.~\ref{Fig_projectors}. (a) Starting point is the original Hamiltonian ${\cal H} = {\cal H}_0 + {\cal H}_1$ with unperturbed eigenvalues $E_n$ and non-zero transitions between 
different eigenstates $|n\rangle$ caused by the interaction part ${\cal H}_1$ (indicated by blue arrows). The corresponding  off-diagonal matrix elements are called  $V_{nm} = \langle n | {\cal H}_1| m \rangle$. (b) Next, suppose the original Hamiltonian 
$\mathcal H$ can be mapped by a unitary transformation \eqref{2.7} to a renormalized Hamiltonian ${\cal H}_\lambda = {\cal H}_{0,\lambda} + {\cal H}_{1,\lambda}$, which only contains  transitions with energy differences smaller than  $\lambda$ (blue shaded area). That is, in $\mathcal H_\lambda$ the particular transitions which combine the states with the highest and lowest energy  (indicated by the red dotted arrows)  have vanished ($V^\lambda_{20}=V^\lambda_{02}=0$).
Here $V_{nm}^\lambda = \langle n_\lambda | {\cal H}_{1,\lambda} | m_\lambda \rangle $ denote the new renormalized off-diagonal matrix elements 
of $\mathcal H_\lambda$. Note that 
due to the unitary transformation all eigenvalues and matrix elements have become renormalized. (c) By further decreasing the cutoff  
$\lambda$ to a new value  $\lambda -\Delta \lambda$ all transitions between $\lambda$ and $\lambda - \Delta \lambda$ will be eliminated. 
In the toy model of Fig.~\ref{Fig_projectors} 
the transitions with the next smaller energy difference are eliminated
($V^{\lambda-\Delta \lambda}_{10}=V^{\lambda-\Delta\lambda}_{01}=0$). (d) In the next steps a sequence of {\it small} transitions 
$\Delta \lambda$ is performed by successively reducing the cutoff until finally $\lambda=0$ is reached. In our particular 
example the remaining transition with the lowest possible non-zero energy difference ($V_{12}$) is eliminated.  
The resulting Hamiltonian  
$\tilde{\mathcal H}$ is fully diagonal and its eigenvalues 
$\tilde{E}_n$ are the same as the ones of the  original Hamiltonian.
}
  \label{Fig_1_PRM}
\end{figure*}

\subsubsection{Renormalization scheme}
\label{II.A.1}

In this subsection the renormalization scheme of our method will be developed on the basis
of transformation \eqref{2.7}. However, instead of a single step, as is formally done in Eq.~\eqref{2.7}, 
a whole sequence of small transformation steps $\Delta \lambda$ will be used.  
Let us consider a step 
from cutoff $\lambda$ to a somewhat reduced cutoff $\lambda - \Delta \lambda$ as illustrated in  Fig.~\ref{Fig_1_PRM}(b,c). 
Its transformation reads
\begin{eqnarray}
  \label{2.8}
  \mathcal{H}_{\lambda-\Delta\lambda} = 
  e^{X_{\lambda, \Delta\lambda}}\; \mathcal{H}_{\lambda}\;  
  e^{-X_{\lambda, \Delta\lambda}} ,
\end{eqnarray}
where $\mathcal{H}_{\lambda}$ and  $\mathcal{H}_{\lambda-\Delta \lambda}$ 
are the Hamiltonians before and  after the step. 
Whereas  $\mathcal{H}_{\lambda}$ is composed of all transitions with excitation energies smaller than $\lambda$, 
 Hamiltonian $\mathcal H_{\lambda - \Delta \lambda}$ 
is the resulting Hamiltonian which only includes transitions with energies smaller than $\lambda -\Delta \lambda$. In Fig.~\ref{Fig_1_PRM}(c) this property is shown by the remaining transition inside the blue shaded area. 
$X_{\lambda, \Delta \lambda}$ is the generator for the unitary transformation step from $\lambda$ 
to $\lambda - \Delta \lambda$.  
Since Eq.~\eqref{2.8} relates Hamiltonian $\mathcal H_\lambda$
to
$\mathcal H_{\lambda-\Delta\lambda}$,  it establishes difference equations or renormalization equations
between the parameters of $\mathcal{H}_{\lambda}$ and $\mathcal H_{\lambda- \Delta \lambda}$. This effect is visualized in Fig.~\ref{Fig_1_PRM}(b,c) accompanied by a slight change of the eigenvalues in the step from (b) to (c).

The solution of the renormalization equations is reached as follows: 
Starting point is the original Hamiltonian ${\cal H}$ which will be called $\mathcal H= \mathcal H_\Lambda$.
Thereby, $\Lambda$ is the maximum cutoff energy  
for transitions  due to $\mathcal H_1$ between the eigenstates of $\mathcal H_0$ (compare Fig.~\ref{Fig_1_PRM}(a)). Next,
 transformation \eqref{2.8} is applied to $\mathcal H_\Lambda$ in order to eliminate all  transitions
between  $\Lambda$ and a slightly reduced cutoff $\Lambda- \Delta \lambda$.
Thereby, Hamiltonian $\mathcal H_\Lambda$ will be renormalized to $\mathcal H_{\Lambda - \Delta \lambda}$.  
 In subsequent small elimination steps $\Delta \lambda$  the cutoff  energy will be further reduced 
 to $\Lambda- 2\Delta \lambda,\Lambda- 3\Delta \lambda, \cdots$ until $\lambda=0$ is reached. 
In this limit, which is the situation in Fig.~\ref{Fig_1_PRM}(d), all transitions from $\mathcal H_{1,\lambda}$
have been integrated out completely. Thus, we arrive at the desired diagonal (or quasi-diagonal)  
result ${\cal H}_{\lambda =0}= {\cal H}_{0, \lambda =0}$. 
Note that ${\cal H}_{\lambda =0}$ depends on the parameters of the original Hamiltonian $\mathcal{H}$. 
They serve  as initial parameter values in the renormalization equations 
for the $\lambda$-dependent parameters of $\mathcal H_\lambda$.   

\subsubsection{Explicit evaluation of step $\lambda$ to $\lambda - \Delta \lambda$}
\label{II.A.2}

Next, transformation \eqref{2.8} has to be evaluated explicitly.
 For a sufficiently small transformation step $\Delta \lambda$, Eq.~\eqref{2.8}
can be expanded in powers of $X_{\lambda, \Delta \lambda}$,
\begin{align}
  \mathcal{H}_{\lambda- \Delta \lambda} &=
  \mathcal{H}_\lambda + \left[ X_{\lambda, \Delta \lambda}, \mathcal{H}_\lambda \right]  +
  \frac{1}{2!}
  \left[ X_{\lambda, \Delta \lambda}, \left[ X_{\lambda, \Delta \lambda}, \mathcal{H}_\lambda \right] \right] 
\nonumber  \\
  &+
  \frac{1}{3!}
  \left[ X_{\lambda, \Delta \lambda}, \left[ X_{\lambda, \Delta \lambda}, 
    \left[ X_{\lambda, \Delta \lambda}, \mathcal{H}_\lambda \right]
  \right] \right] + \dots .
   \label{2.9}
  \end{align}
Note that the correct size dependence of the effective Hamiltonian is automatically guaranteed 
by the commutators appearing in Eq.~\eqref{2.9}. 
In order to construct the generator $X_{\lambda, \Delta \lambda}$, we assume that it can be 
presented as a power series in $\mathcal H_{1,\lambda}$,
\begin{equation}
  \label{2.10}
  X_{\lambda,\Delta \lambda} =
  X_{\lambda,\Delta \lambda}^{(1)} + X_{\lambda,\Delta \lambda}^{(2)} + X_{\lambda,\Delta \lambda}^{(3)} +
  \dots 
\end{equation}
with $X_{\lambda,\Delta \lambda}^{(n)} \sim \mathcal O(\mathcal H_{1,\lambda})^n$.
Therefore,  
$\mathcal{H}_{\lambda - \Delta \lambda}$ can be formulated as power series in $\mathcal H_{1,\lambda}$,
\begin{equation}
\begin{split}
  \label{2.11}
  \mathcal{H}_{\lambda- \Delta \lambda} &=
  \mathcal{H}_{0,\lambda} + \mathcal{H}_{1,\lambda} + 
  \left[ X_{\lambda,\Delta \lambda}^{(1)}, \mathcal{H}_{0,\lambda} \right]    \\
 &+ \left[ X_{\lambda,\Delta \lambda}^{(1)}, \mathcal{H}_{1,\lambda} \right]  
  + 
  \left[ X_{\lambda,\Delta \lambda}^{(2)}, \mathcal{H}_{0,\lambda} \right] \\
  &+ \frac{1}{2!} \left[
    X_{\lambda,\Delta \lambda}^{(1)}, \left[ X_{\lambda,\Delta \lambda}^{(1)}, \mathcal{H}_{0,\lambda} \right]
  \right] 
   + {\cal O}(\mathcal{H}_{1,\lambda}^{3}) .
   \end{split}
\end{equation}
The first commutator stands for renormalization contributions 
to first order in $\mathcal H_{1,\lambda}$, whereas the three successive commutators are contributions to
second order in $\mathcal H_{1,\lambda}$. 
Applying relation \eqref{2.6}  (with $\lambda$ replaced by $\lambda -\Delta \lambda$), 
\begin{eqnarray}
\label{2.12}
\mathbf Q_{\lambda-\Delta \lambda} \mathcal H_{\lambda -\Delta \lambda}=0  ,
\end{eqnarray}
the following expressions for $X_{\lambda, \Delta \lambda}^{(n)}$ can successively be deduced, 
 \begin{eqnarray}
  \label{2.13}
  && \mathbf{Q}_{\lambda-\Delta \lambda}^{} X_{\lambda, \Delta \lambda}^{(1)} =
  \frac{1}{{\bf L}_{0,\lambda}} 
  \left( \mathbf{Q}_{\lambda-\Delta \lambda}^{} \mathcal{H}_{1,\lambda} \right),  \\
  &&  \nonumber \\
  \label{2.14}
  &&\mathbf{Q}_{\lambda-\Delta \lambda}^{} X_{\lambda,\Delta \lambda}^{(2)} = \\
 && \quad  -  \frac{1}{2 \mathbf{L}_{0,\lambda} } \mathbf{Q}_{\lambda-\Delta \lambda}^{}  \Big[
    ( \mathbf{Q}_{\lambda-\Delta \lambda}^{}\mathcal{H}_{1,\lambda} ), 
    \frac{1}{\mathbf{L}_{0,\lambda}}
    ( \mathbf{Q}_{\lambda-\Delta \lambda}^{}\mathcal{H}_{1,\lambda})
  \Big]   \nonumber \\
  &&  \quad -
  \frac{1}{{\bf L}_{0,\lambda}}
  \mathbf{Q}_{\lambda-\Delta \lambda}^{}
  \Big[
    (\mathbf{P}_{\lambda-\Delta \lambda}^{}\mathcal{H}_{1,\lambda}), 
    \frac{1}{ \mathbf{L}_{0,\lambda} }
    (\mathbf{Q}_{\lambda-\Delta \lambda}^{}\mathcal{H}_{1,\lambda})
  \Big] ,  \nonumber \\
  && \nonumber \\
&& \mathbf{Q}_{\lambda-\Delta \lambda}^{} X_{\lambda, \Delta \lambda}^{(3)} = \cdots . \nonumber 
\end{eqnarray}
Here the quantity ${\bf L}_{0,\lambda}$ is the so-called Liouville operator belonging to the unperturbed
Hamiltonian $\mathcal{H}_{0,\lambda}$. It  is defined by 
$\mathbf{L}_{0,\lambda}\mathcal{A} = [\mathcal{H}_{0,\lambda}, \mathcal{A}]$
for any operator variable $\mathcal{A}$. Relations \eqref{2.13}, \eqref{2.14}, ... 
make a statement  about the high energy parts 
$\mathbf Q_{\lambda -\Delta \lambda} X^{(n)}_{\lambda, \Delta \lambda}$
of $X^{(n)}_{\lambda, \Delta \lambda}$, which are composed of transitions within the interval $\Delta \lambda$ with energies  between $\lambda- \Delta \lambda$ and $\lambda$.  
Note that the parts  $\mathbf P_{\lambda -\Delta \lambda} X^{(n)}_{\lambda, \Delta \lambda}$ of
$X^{(n)}_{\lambda, \Delta \lambda}$ with excitation energies below $\lambda -\Delta \lambda$ 
are not fixed by Eq.~\eqref{2.12}. 

Due to this freedom of choice it is natural to decompose the generator $X_{\lambda, \Delta \lambda}$  for the renormalization step from $\lambda$ to $\lambda- \Delta \lambda$ in a low-energy part
${\mathbf P}_{\lambda - \Delta \lambda} X_{\lambda, \Delta \lambda}$ and a high-energy part 
${\mathbf Q}_{\lambda - \Delta \lambda} X_{\lambda, \Delta \lambda}$ according to,  
\begin{equation*}
  \label{A.1}
  X_{\lambda, \Delta \lambda} =
  {\mathbf P}_{\lambda - \Delta \lambda} X_{\lambda, \Delta \lambda} + 
  {\mathbf Q}_{\lambda - \Delta \lambda} X_{\lambda, \Delta \lambda} .
\end{equation*} 
Due to construction, in 
$ {\mathbf P}_{\lambda - \Delta \lambda} X_{\lambda, \Delta \lambda}$ only
excitations are included with energies smaller than $\lambda- \Delta \lambda$, whereas 
${\mathbf Q}_{\lambda - \Delta \lambda} X_{\lambda, \Delta \lambda}$ only contains 
 transitions with energies between $\lambda - \Delta \lambda$ and $\lambda$. 
 Thereby, ${\mathbf Q}_{\lambda - \Delta \lambda} X_{\lambda, \Delta \lambda}$
ensures that condition \eqref{2.12} is fulfilled. Its series expansion must fulfill Eqs.~\eqref{2.13},\eqref{2.14}, $\dots$, so that  
there is no freedom of choice for this part of the generator.

\begin{figure}[h]
  \centering
\includegraphics[width=0.9\columnwidth]{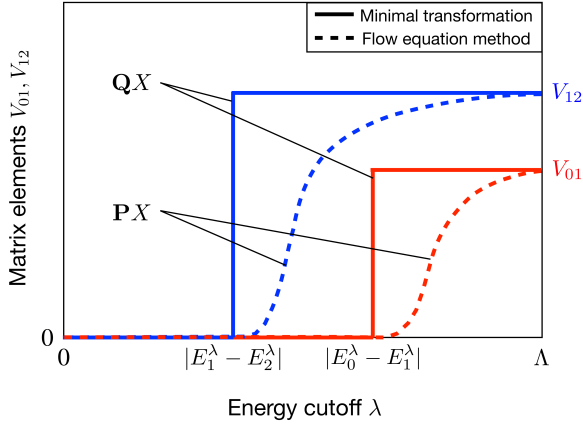}
  \caption{
 (Color online) Schematic behavior of some interaction matrix elements  $V^\lambda_{nm}$ as a function of $\lambda$ for an infinitely  large number of small renormalization steps ($\Delta\lambda \rightarrow 0$) for the two choices of ${\mathbf P}_{\lambda - \Delta \lambda} X_{\lambda, \Delta \lambda}$ discussed in the text. 
  In Wegner's flow equation method  (dashed lines) the part $\mathbf{P}X$ is chosen such that the matrix elements $V^\lambda_{nm}$ continuously flow to zero {\it before} $\lambda$ reaches their  respective transition energies $|E^\lambda_n -E^\lambda_m|$. 
  In this case the part $\mathbf{Q}X$ can be neglected.  In the minimal transformation (solid lines) the reverse situation is found. 
  Here all matrix elements $V^\lambda_{nm}$ remain constant for all steps down to their particular $\lambda$-values 
  where they suddenly drop to zero. The $\lambda$-values where this happens are fixed by their respective  transition energies (compare Figs.~\ref{Fig_projectors} and \ref{Fig_1_PRM}). That is, in the minimal transformation the 'high-energy' part $\mathbf{Q}X$ is solely responsible for the elimination of the matrix elements.  
  }
  \label{Fig_matrix_el}
\end{figure}

On the other hand, the part of low energy transitions, ${\mathbf P}_{\lambda - \Delta \lambda} X_{\lambda, \Delta \lambda}$, is not fixed. 
As is shown below for a particular example the  renormalization of the {\it off-diagonal} matrix elements $V_{nm}^\lambda$ of 
${\cal H}_\lambda$ depends on a particular choice of  ${\mathbf P}_{\lambda - \Delta \lambda} X_{\lambda, \Delta \lambda}$ 
(compare Fig.~\ref{Fig_1_PRM}(b,c) and Fig.~\ref{Fig_matrix_el}). In practice, however, only two specific  choices have become established so far. The difference is visualized in Fig.~\ref{Fig_matrix_el} for two matrix elements of the example system considered above.    One possibility (shown by the dashed lines in Fig.~\ref{Fig_matrix_el}) is to fix a particular expression for ${\mathbf P}_{\lambda - \Delta \lambda} X_{\lambda, \Delta \lambda}$ such that almost all low-energy transitions are already 
integrated out  \textit{before} a small cutoff 
energy $\lambda$ is reached. In this case the influence of the high-energy part 
${\mathbf Q}_{\lambda - \Delta \lambda} X_{\lambda, \Delta \lambda}$
becomes small. As is also shown below for a simple model system this particular choice  
leads to Wegner's flow equation method \cite{W_1994}
as well as  the similarity transformation by G\l acek and Wilson \cite{GW_1993, GW_1994}.  

Another reasonable choice of the low-energy part is to simply  set it to zero,
\begin{equation}
 \label{2.15}
  \mathbf P_{\lambda-\Delta \lambda} X_{\lambda,\Delta \lambda} = 
 \mathbf P_{\lambda-\Delta \lambda} X_{\lambda,\Delta \lambda}^{(n)}
  = 0 \, ,
\end{equation} 
which is a choice that will be called 'minimal' transformation. This situation is shown by the solid lines in Fig.~\ref{Fig_matrix_el}. It leads to the effect that each matrix element is kept constant until a particular $\lambda$ value is reached. At this cutoff the corresponding  matrix element is suddenly  eliminated by the orthogonal part 
$\mathbf Q_{\lambda-\Delta \lambda} X_{\lambda,\Delta \lambda}$ of the generator.  The particular choice \eqref{2.15} allows to  derive an explicit expression for the effective Hamiltonian ${\cal H}_{\lambda- \Delta \lambda}$ 
at the reduced cutoff $\lambda -\Delta \lambda$ 
from  the  former Hamiltonian ${\cal H}_{\lambda}$ at cutoff $\lambda$. The corresponding renormalization scheme is also known under the name projector-based renormalization method (PRM) which was introduced in Ref.~\cite{BHS_2002}. 

Inserting Eqs.~\eqref{2.13}-\eqref{2.15}  into Eq.~\eqref{2.11} one finds the corresponding series expansion with respect to ${\cal H}_{1,\lambda}$,
\begin{align}
\label{2.16}
  {\cal H}_{\lambda- \Delta \lambda} &=
  {\cal H}_{0,\lambda} + \mathbf P_{\lambda- \Delta \lambda}^{}{\cal H}_{1,\lambda} \nonumber \\
   &- \frac{1}{2} \mathbf P_{\lambda- \Delta \lambda}^{}
  \left[
    (\mathbf Q_{\lambda- \Delta \lambda}^{}{\cal H}_{1,\lambda}), \frac{1}{\mathbf{L}_{0,\lambda}}
    (\mathbf Q_{\lambda- \Delta \lambda}^{}{\cal H}_{1,\lambda})
  \right] \nonumber\\
   &-
  \mathbf P_{\lambda -\Delta \lambda}^{}
  \left[
    (\mathbf P_{\lambda- \Delta \lambda}^{}{\cal H}_{1,\lambda}), \frac{1}{ \mathbf{L}_{0,\lambda} }
    (\mathbf Q_{\lambda- \Delta \lambda}^{}{\cal H}_{1,\lambda})
  \right] \nonumber \\
  &+ \mathcal{O}(\mathcal{H}_{1,\lambda}^{3})  .
\end{align}
 This result may also be derived in a slightly different way. According to requirement 
 ${\mathcal H}_{\lambda- \Delta \lambda} = 
 {\bf P}_{\lambda - \Delta \lambda} {\mathcal H}_{\lambda-\Delta \lambda}$  one first deduces 
 from Eq.~\eqref{2.11} 
 \begin{align}
  \label{2.17}
  \mathcal{H}_{\lambda- \Delta \lambda} &=
  \mathcal{H}_{0,\lambda} + \mathbf P_{\lambda- \Delta \lambda} \mathcal{H}_{1,\lambda} + 
  \mathbf P_{\lambda- \Delta \lambda} \left[ X_{\lambda,\Delta \lambda}^{(1)}, \mathcal{H}_{0,\lambda} \right]  \nonumber \\
  &+ \mathbf P_{\lambda- \Delta \lambda}\left[ X_{\lambda,\Delta \lambda}^{(2)}, \mathcal{H}_{0,\lambda} \right]  + \mathbf P_{\lambda- \Delta \lambda}\left[ X_{\lambda,\Delta \lambda}^{(1)}, \mathcal{H}_{1,\lambda} \right] \nonumber \\
  &+ \frac{1}{2!} \mathbf P_{\lambda- \Delta \lambda} \left[
    X_{\lambda,\Delta \lambda}^{(1)}, \left[ X_{\lambda,\Delta \lambda}^{(1)}, \mathcal{H}_{0,\lambda} \right]
  \right] + {\cal O}(\mathcal{H}_{1,\lambda}^{3}) , 
\end{align}
where the second and third commutator drop due to Eqs.~\eqref{2.13}  and \eqref{2.14}.
Thus, using Eq.~\eqref{2.15} one finds  
\begin{align}
\label{2.18}
 {\cal H}_{\lambda- \Delta \lambda} &=
  {\cal H}_{0,\lambda} + \mathbf P_{\lambda- \Delta \lambda}^{}{\cal H}_{1,\lambda}   
+ \mathbf P_{\lambda- \Delta \lambda}\left[X^{(1)}_{\lambda, \Delta \lambda}, \mathcal H_{1,\lambda}\right] \nonumber \\
&- \frac{1}{2} \mathbf P_{\lambda- \Delta \lambda}\left[X^{(1)}_{\lambda, \Delta \lambda}, 
{\bf Q}_{\lambda-\Delta \lambda} \mathcal H_{1,\lambda} \right] 
+ \mathcal O (\mathcal H^3_{1,\lambda})
, 
\end{align}
which with Eq.~\eqref{2.13} immediately leads back to result \eqref{2.16}. 
Since $X_{\lambda, \Delta \lambda}^{(2)}$ dropped in Eq.~\eqref{2.18}, its only task in Eq.~\eqref{2.11} 
is to fulfill requirement  \eqref{2.12}.

Expressions \eqref{2.16} or \eqref{2.18} 
represent the desired relation between $\mathcal H_\lambda$ and 
$\mathcal H_{\lambda - \Delta \lambda}$ with renormalization contributions up to second order 
in $\mathcal H_{1,\lambda}$.  As aforementioned, the complete renormalization 
scheme is based on a whole sequence of small unitary renormalization steps $\Delta \lambda$ between
$\lambda =\Lambda$ and $\lambda=0$.

An alternative, yet  approximate formulation 
for the renormalization step from $\lambda$ to $\lambda- \Delta \lambda$ 
starts from equation \eqref{2.17} (where $X^{(2)}_{\lambda, \Delta \lambda}$
has dropped). Replacing for a moment the name of the first order generator $X^{(1)}_{\lambda,\Delta \lambda}$ by $X_{\lambda, \Delta \lambda}$, 
one arrives at 
\begin{align*}
  \label{2.19}
  \mathcal{H}_{\lambda- \Delta \lambda} &\approx
  \mathcal{H}_{0,\lambda} + \mathbf P_{\lambda- \Delta \lambda} \mathcal{H}_{1,\lambda} + 
  \mathbf P_{\lambda- \Delta \lambda} \left[ X_{\lambda,\Delta \lambda}, \mathcal{H}_{0,\lambda} \right] \nonumber \\ 
  &+ \mathbf P_{\lambda- \Delta \lambda}\left[ X_{\lambda,\Delta \lambda}, \mathcal{H}_{1,\lambda} \right]  \nonumber \\
  &+ \frac{1}{2!} \mathbf P_{\lambda- \Delta \lambda} \left[
    X_{\lambda,\Delta \lambda}, \left[ X_{\lambda,\Delta \lambda}, \mathcal{H}_{0,\lambda} \right]
  \right] ,
\end{align*}
where  higher order terms from expansion \eqref{2.17} have been neglected.  
Now we assume  the generator $X_{\lambda, \Delta \lambda}$ to be non-perturbative and make
an {\it ansatz } for the generator, which  has the same operator structure 
 as $X_{\lambda,\Delta \lambda}^{(1)}$.  Often it turns out that this procedure is a  good choice. 
This strategy has  been applied successfully to a number of problems such as the periodic Anderson model 
or the Holstein model \cite{HB_2005,SHBWF_2005}.
In particular, in this way possible divergent contributions from the perturbative renormalization treatment can be avoided  (see below).

\subsubsection{Evaluation of expectation values}

To study physical quantities of many-particle systems also  expectation values have to be
evaluated. For instance, the expectation value of an operator
variable $\mathcal A$ in thermal equilibrium $\langle \mathcal A \rangle =  \mathrm{Tr}( \mathcal{A} \, e^{-\beta\mathcal{H}})/\mathrm{Tr} \,e^{-\beta\mathcal{H}}$
can be rewritten by exploiting the invariance against unitary transformations of operator expressions under a trace,  
\begin{equation}
\label{2.21}
  \langle \mathcal{A} \rangle =
  \frac{
    \mathrm{Tr}\left( {\mathcal{A}_\lambda} 
\,e^{-\beta{\mathcal{H}_\lambda}} \right) }
  { \mathrm{Tr} \,e^{-\beta{\mathcal{H}_\lambda}} }
\end{equation}
with $\mathcal A_\lambda = e^{X_\lambda} \mathcal A e^{-X_\lambda}$. Thereby, ${X_\lambda}$ is
again a compact notation for the generator combining the initial cutoff $\Lambda$ and $\lambda$. 
For $\lambda \rightarrow 0$ one obtains
\begin{equation*}
\label{2.22}
  \langle \mathcal{A} \rangle =  \frac{
    \mathrm{Tr}\left( \tilde{\mathcal{A}} 
\,e^{-\beta\tilde{\mathcal{H}}} \right) }   { \mathrm{Tr} \,e^{-\beta\tilde{\mathcal{H}}} } 
  =: \langle \tilde {\mathcal A} \rangle_{\tilde {\mathcal H}}, 
\end{equation*}
with 
$\tilde{\mathcal{A}} = \lim_{\lambda \rightarrow 0}
\mathcal{A}_{\lambda}$.  
Note that not only $\mathcal H$ but also the operator variable $\mathcal A$  is subject to the same unitary
transformation. Assuming the operator $\mathcal A_\lambda$ at cutoff $\lambda$ is known, the renormalized 
operator $\mathcal A_{\lambda -\Delta \lambda}$ after a small renormalization step $\Delta \lambda$
reads
\begin{align}
\label{2.23}
\mathcal A_{\lambda - \Delta \lambda} &= \mathcal A_{\lambda} + [ X_{\lambda, \Delta \lambda}, 
\mathcal A_{\lambda}] \nonumber \\
&+  \frac{1}{2}   [ X_{\lambda, \Delta \lambda},   [ X_{\lambda, \Delta \lambda},\mathcal A_{\lambda}] ] 
+\cdots .
\end{align}
Relation \eqref{2.23} is used to derive additional renormalization equations for $\mathcal{A}_{\lambda}$ in analogy to those for $\mathcal{H}_\lambda$. 

Sometimes it is favorable  to determine
expectation values from the free energy $F$, from which 
expectation values are obtained by functional derivatives. Since
$\tilde{\mathcal{H}}$ and $\mathcal H$ are unitarily connected, one has 
\begin{equation*}
\label{2.24}
  F = 
  - \frac{1}{\beta} \mathrm{ln} \, \mathrm{Tr} \, e^{-\beta\mathcal{H}}
  \,=\,
  - \frac{1}{\beta} \mathrm{ln} \, \mathrm{Tr} \, 
e^{-\beta\tilde{\mathcal{H}}} .
\end{equation*}
Thus, $F$ may be easily evaluated from the diagonal (or quasi-diagonal)
$\tilde{\mathcal{H}}$. Examples are found in 
Refs.~\cite{HB_2005} and \cite{HB_2003}.

\subsection{Exactly solvable model}
\label{App A1}

Let us now illustrate the concepts of the generalized diagonalization scheme on the basis of an exactly solvable model. For this purpose we consider a specific Hamiltonian ${\mathcal H}$ describing a system of two types of  spinless fermions which can hybridize with each other. Such a model, written in the decomposition ${\mathcal H} = {\cal H}_{0} + {\cal H}_{1}$, may read
\begin{equation}
\begin{split}
 {\mathcal H}_{0} &=
  \sum_k
  \left(
    \varepsilon_{f} \, f^{\dagger}_{k} f_{k}^{} +
    \varepsilon_{k} \, c^{\dagger}_{k} c_{k}^{}
  \right) ,  \\
  %
  {\cal H}_{1} &=
  \sum_k V_{k}
  \left(
    f_{k}^{\dagger}c_{k}^{} + c_{k}^{\dagger}f_{k}^{}
  \right).  \label{A.2}
  \end{split}
\end{equation}
The index $k$ denotes wave numbers, and the one-particle energies
$\varepsilon_f$ and $\varepsilon_k$ are measured with respect to the chemical potential. The hybridization strength for a particular wave number  is described by the parameter $V_{k}$. The quadratic form of the fermion operators and the lack of interaction between the fermions makes the model particularly simple and exactly solvable. The model  is known in the context of the so-called Fano-Anderson model \cite{A_1961, F_1961,HB_2005} which has been introduced for a simplified description of  dispersionless $f$-electrons which hybridize with  conduction electrons.

Usually our method is constructed to integrate out an interaction term of the Hamiltonian. In the specific case of model \eqref{A.2} the hybridization term is considered as ${\cal H}_{1}$ instead. In this way the coupling between the two fermions will  be integrated out instead of an interaction  which leads to two independent systems of  renormalized fermions. The result will be compared with the exact  diagonalization  of the Hamiltonian using a rotation of the Hilbert space of fermions. Purpose of these  considerations is  only  to demonstrate the general idea 
of the renormalization scheme and the introduced concepts of the  low energy generator  
part $\mathbf P_{\lambda-\Delta \lambda} X_{\lambda, \Delta \lambda}$ in terms of a simplest possible model. 
In this way the role of the continuous flow equation renormalization  \cite{W_1994} within our generalized diagonalization method 
may become more clear. Note that a true interaction between the two types of fermions in the model \eqref{A.2} is also considered in Subsec.~\ref{III.B}.

At first, the model \eqref{A.2} can easily be diagonalized,
\begin{equation*}
  \label{A.3}
  {\cal H} =
  \sum_{k} \omega_{k}^\alpha
  \alpha_{k}^{\dagger} \, \alpha_{k}^{} +
  \sum_{k} \omega_{k}^\beta \,
  \beta_{k}^{\dagger} \beta_{k}^{},
\end{equation*}
where the eigenmodes $\alpha_{k}^{\dagger}$ and $\beta_{k}^{\dagger}$ are 
linear combinations of the original fermionic operators 
$c_{k}^{\dagger}$ and $f_{k}^{\dagger}$,
\begin{equation}
\label{B21}
  \alpha_{k}^{\dagger} =
  u_{k} \, f_{k}^{\dagger} + v_{k} \,
  c_{k}^{\dagger}, \quad \beta_{k}^{\dagger} =
  -v_{k} \, f_{k}^{\dagger} + u_{k} \,
  c_{k}^{\dagger}, 
\end{equation}
with
\begin{equation}
\label{A.5}
  |u_{k}|^{2} =
  \frac{1}{2}
  \left( 1 - \frac{\varepsilon_{k}-\varepsilon_{f}}
  {W_{k}} \right), \quad
  |v_{k}|^{2} =
  \frac{1}{2}
  \left( 1 + \frac{\varepsilon_{k}-\varepsilon_{f}}
{W_{k}} \right).
\end{equation}
The quantity $ W_{k}$ in Eqs.~\eqref{A.5} is defined by
$ W_{k} = 
  [ \left( \varepsilon_{k}-\varepsilon_{f} \right)^{2} +
    4 |V_{k}|^{2}  ]^{(1/2)}$, and the eigenvalues $\omega_{k}^\alpha$ and $\omega_{k}^\beta$ of $\mathcal{H}$ are given by
\begin{equation}
  \label{A.6}
  \omega_{k}^{\alpha,\beta} =
  \frac{\varepsilon_{k} + \varepsilon_f}{2}
  \pm \frac{W_k}{2}.
\end{equation}
In Fig.~\ref{Fig_Fano}(a) the two eigenvalues are shown as a function of $k$ for an actual numerical example with linear dispersions  $\varepsilon_k = \Lambda (|k| - 1)$, $\varepsilon_f = 0$ (dotted lines) and constant hybridization strength $V_k = V$. The two eigenvalues $\omega_{k}^{\alpha,\beta}$ form two $k$-dependent branches (combined blue and red solid lines) which are distinct in energy. They describe the typical hybridizatation gap known from some heavy fermion materials.  

\subsubsection{Minimal transformation approach}
\label{App A1a}

\begin{figure}[h]
  \centering
\includegraphics[width=\columnwidth]{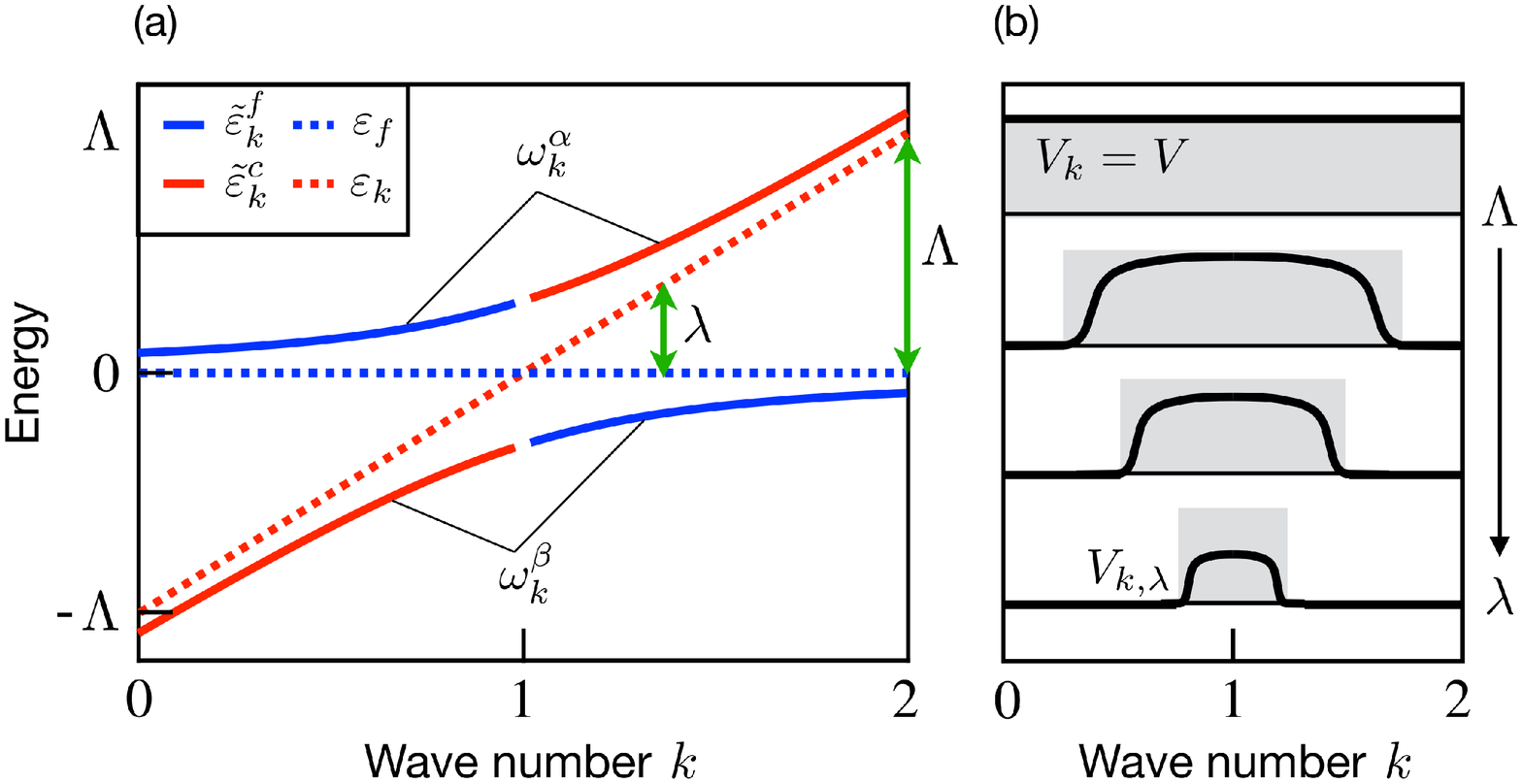}
  \caption{
    (Color online) (a) Fully renormalized branches of the exactly solvable fermion model \eqref{A.2} both for the minimal transformation approach and the flow equation method. The solid lines in blue and red indicate the branches for the 
    '$f$' type and '$c$' type electrons. The dispersions of the original model $\mathcal H_0$ are also shown (dotted lines).  
  Note that within our method the branches are identical and keep their characters from the original dispersions (blue and red, respectively) accompanied by a discontinuity at the original crossing point.  In contrast, the two branches \eqref{A.6} from the exact diagonalization  change their character from '$f$' to '$c$' type (from blue to red) and vice versa at the crossing point. 
     (b) Coupling parameter $V_{k,\lambda}$ as a function of $k$ for four different $\lambda$ values starting from the largest possible value $\Lambda$ to a rather small value (compare corresponding transitions marked by green arrows in panel (a)). The grey shaded areas indicate the minimal transformation (with $V_{k,\lambda}= V_k \Theta_{k, \lambda}$) with a reduced extension for decreasing $\lambda$.  Within the flow equation method a continuous shrinking  is found (solid lines). 
  }
  \label{Fig_Fano}
\end{figure}

 The first step is to formulate an appropriate {\it ansatz} for the renormalized Hamiltonian ${\cal H}_{\lambda}$ which we have introduced in Sec.~\ref{II.A}. Its simplest possible formulation  consists   of  the same operator structure as the original Hamiltonian ${\cal H}$, i.~e.~only the energy  parameters of ${\cal H}_{\lambda}$ become $\lambda$-dependent whereas all operators remain unchanged. Transferring this idea to the present system an appropriate {\it ansatz} ${\cal H}_{\lambda} = 
  {\cal H}_{0,\lambda} + {\cal H}_{1,\lambda}$ reads as follows,
\begin{equation}
  \label{A.7}
  \begin{split}
{\cal H}_{0, \lambda} &=
   \sum_{k}
  \left(
    \varepsilon_{k,\lambda}^{f} \, f^{\dagger}_{k} f_{k}^{} +
    \varepsilon_{k,\lambda}^{c} \, c^{\dagger}_{k} c_{k}^{}
  \right),  \\
{\cal H}_{1,\lambda} &=
  \sum_{k}   V_{k} \Theta_{k, \lambda}
  \left(
    f_{k}^{\dagger}c_{k}^{} + c_{k}^{\dagger}f_{k}^{}
  \right) ,
 \end{split}
\end{equation}
where ${\cal H}_{1,\lambda}$  includes a cutoff function $\Theta_{k, \lambda}= 
\Theta( \lambda -|\varepsilon^f_{k,\lambda} - \varepsilon^c_{k,\lambda} | )$
in order to ensure the requirement
$\mathbf{P}_{\lambda}\mathcal{H}_{\lambda}=\mathcal{H}_{\lambda}$. Note that the operator structure in \eqref{A.7} is kept fixed and the $\lambda$-dependence is transferred to  the  parameters. Furthermore note that the 'minimal' transformation ${\bf P}_{\lambda- \Delta \lambda} X_{\lambda,\Delta\lambda} =0$  will be used. As seen in Fig.~\ref{Fig_matrix_el} within this concept the coupling matrix element $V_k$ is $\lambda$-independent. However, $V_{k,\lambda}=V_k \Theta_{k,\lambda}$  abruptly drops to zero at the particular $\lambda$ value which is equal to the corresponding energy difference. This is  described by the presence of the cutoff function $ \Theta_{k, \lambda}$ in ${\cal H}_{1,\lambda}$ and visualized in Fig.~\ref{Fig_Fano}(b) by the grey shaded areas.

In the next step we eliminate excitations with energies 
between $\lambda$ and $\lambda - \Delta\lambda$ by means of the
unitary transformation  \eqref{2.8}. By inspection of the perturbation
expansion \eqref{2.10} for the present model, the generator $X_{\lambda, \Delta \lambda}$ of
the unitary transformation must have the general operator form:
\begin{equation}
 \label{A.8}
  X_{\lambda,\Delta\lambda} =
  \sum_{k}
  A_{k}(\lambda,\Delta\lambda)
  \left(
    f_{k}^{\dagger} c_{k}^{} - 
    c_{k}^{\dagger} f_{k}^{}
  \right),  
\end{equation}
where the 'minimal' transformation ${\bf P}_{\lambda- \Delta \lambda} X_{\lambda,\Delta\lambda} =0$  was used.
The yet unknown coefficients $A_{k}(\lambda,\Delta\lambda)$ are found
from condition \eqref{2.12}. Starting point is  transformation \eqref{2.8}, which must be applied to 
Hamiltonian \eqref{A.7}.  This leads for instance to
\begin{eqnarray}
\label{A.9}
  \lefteqn{
    e^{X_{\lambda,\Delta\lambda}} \, 
    c^{\dagger}_{k}c_{k}^{} \,
    e^{-X_{\lambda,\Delta\lambda}} - c^{\dagger}_{k}c_{k}^{}
    \,=\,
  }&&    \nonumber \\
  &=&
  \frac{1}{2}
  \left\{\cos\left[ 2 A_{k}(\lambda,\Delta\lambda)\right] - 1 \right\}
  \left( 
    c^{\dagger}_{k}c_{k}^{} - 
    f^{\dagger}_{k}f_{k}^{} 
  \right)  \nonumber \\
  && +\,
  \frac{1}{2}
  \sin\left[ 2 A_{k}(\lambda,\Delta\lambda)\right]
  \left(
    f^{\dagger}_{k}c_{k}^{} + 
    c^{\dagger}_{k}f_{k}^{} 
  \right)
  \nonumber
\end{eqnarray}
and to  similar expressions for $f^{\dagger}_{k}f_{k}^{}$ and 
$(  f^{\dagger}_{k}c_{k}^{} + 
  c^{\dagger}_{k}f_{k}^{} )$. 
Note that different $k$ values do not couple with each other.
Inserting the above transformations into Eq.~\eqref{2.8} and comparing the result with Eq.~\eqref{A.7} considered at cutoff $\lambda - \Delta\lambda$  the
following renormalization equations are found, 
\begin{eqnarray}
  \label{A.10}
  \lefteqn{
    \varepsilon^{f}_{k, \lambda-\Delta\lambda} - 
    \varepsilon^{f}_{k,\lambda} \,= V_{k,\lambda} \sin\left[2 A_{k}
(\lambda,\Delta\lambda)\right]
  }&&\\
  &-&
  \frac{1}{2}
  \left\{
    \cos\left[ 2 A_{k}(\lambda,\Delta\lambda)\right] - 1
  \right\}
  \left( 
    \varepsilon^{c}_{k,\lambda} -
    \varepsilon^{f}_{k,\lambda}
  \right), \nonumber \\
  \label{A.11}
  \lefteqn{ 
  \varepsilon^{c}_{k,\lambda-\Delta\lambda} - 
  \varepsilon^{c}_{k,\lambda} 
  \,=\,
  -\, \left(
    \varepsilon^{f}_{k,\lambda-\Delta\lambda} - 
    \varepsilon^{f}_{k,\lambda}
  \right).}&&
\end{eqnarray}
To determine the coefficients $A_{k}(\lambda,\Delta\lambda)$ we employ 
condition $\mathbf{Q}_{\lambda-\Delta\lambda}\mathcal{H}_{\lambda-\Delta\lambda}=0$ [Eq.~\eqref{2.12}]. 
Taking moreover the low excitation-energy part
of the generator equal to zero,  
$\mathbf{P}_{\lambda-\Delta\lambda}X_{\lambda,\Delta\lambda}=0$, we find
\begin{equation}
  \label{A.12}
    \tan\left[ 2 A_{k}(\lambda,\Delta\lambda) \right] 
  =
    \left[1 - \Theta_{k,\lambda- \Delta \lambda } \right] 
 \Theta_{k,\lambda} \,
  \frac{2V_{k}}
  {\varepsilon^{f}_{k,\lambda} -
    \varepsilon^{c}_{k,\lambda}} . 
\end{equation}
Result \eqref{A.12} shows that also $A_k(\lambda, \Delta \lambda)$ contains the 
cutoff factors $  \left[1 - \Theta_{k,\lambda- \Delta \lambda } \right] \Theta_{k,\lambda}$. Furthermore, the following relation
$
  |\varepsilon^{f}_{k,\lambda} -
  \varepsilon^{c}_{k,\lambda}| \le
  |\varepsilon^{f}_{k,\lambda-\Delta\lambda} -
  \varepsilon^{c}_{k,\lambda-\Delta\lambda}|
$ is fulfilled.
Thus, each $k$ value is renormalized only once during the
renormalization procedure which eliminates excitations from large to small
$\lambda$ values. For such a steplike renormalization it is easy to sum up all renormalization 
steps between the original cutoff $\lambda=\Lambda$ and $\lambda=0$. 
Thus, replacing $\lambda$ by $\Lambda$ and setting $\lambda- \Delta \lambda=0$  
in Eqs.~\eqref{A.10} -\eqref{A.12}
one immediately finds for the fully renormalized Hamiltonian $\tilde{\mathcal{H}} := 
  \lim_{\lambda\rightarrow 0} \mathcal{H}_{\lambda} =
  \sum_{k} (
    \tilde{\varepsilon}_{k}^{f}\,
    f^{\dagger}_{k} f_{k}^{} + 
    \tilde{\varepsilon}_{k}^{c}\,
    c^{\dagger}_{k} c_{k}^{})$
which is diagonal. The renormalized energies are given by
\begin{equation}
  \label{A.14}
  \tilde{\varepsilon}_{k}^{(f,c)} =
  \frac{\varepsilon_{f} +\varepsilon_{k}}{2} \pm
   \mathrm{sgn}( \varepsilon_{f} -\varepsilon_{k} )
  \frac{
    W_{k} 
  }{2}
 .
\end{equation}
 Here we have taken into account that $A_{k}(\lambda, \Delta \lambda)$ changes its sign if the difference 
$\varepsilon^f_{k,\lambda} - \varepsilon^c_{k,\lambda}$ changes its sign.
Note that the eigenvalues $\tilde{\varepsilon}_{k}^{f}$ and $\tilde{\varepsilon}_{k}^{c}$
from our renormalization approach correspond to the eigenvalues $\omega_{k}^{(\alpha, \beta)}$ from the 
exact diagonalization  [Eq.~\eqref{A.6}].  
However, there is an important difference between the two
approaches: In the exact diagonalization  the eigenenergies $\omega_{k}^{(\alpha, \beta)}$  
and the eigenmodes $\alpha_{k}^{\dagger}$ and
$\beta_{k}^{\dagger}$ change their character as a function of wave vector $k$, i.~e.~they are either $f$- or $c$-like depending on the sign of ($\varepsilon_{k} - \varepsilon_f$) [compare Eq.~\eqref{A.6}].
 In contrast,  in our approach the eigenenergies $\tilde \varepsilon^f_{k}$  and 
 $\tilde \varepsilon^c_{k}$ as well as the eigenmodes always keep their own $f$- or $c$-character.
 This feature becomes manifest in the terms $[{\rm sgn}\big(\varepsilon_f -  \varepsilon_{k}\big) (W_{k}/2)]$ 
  in Eq.~\eqref{A.14} and is visualized in Fig.~\ref{Fig_Fano}(a) by the two different colors of the solid lines. 
In particular, the
quasi-particle energies $\tilde{\varepsilon}_{k}^{f}$ and
$\tilde{\varepsilon}_{k}^{c}$ show a steplike behavior as a function of
$k$ at $\varepsilon_{f} = \varepsilon_{k}$ (crossing point of the dotted lines). 
Thereby, the deviations from the original one-particle energies $\varepsilon_{f}$ and
$\varepsilon_{k}$ remain relatively small for all $k$ values.
Moreover, the renormalization contributions in our method have to be summed up  to all orders in the 'perturbation'. Only then complete agreement with the exact diagonalization is achieved. However, for realistic many-particle systems 
with 'true' many-particle interactions this complication of quadratic terms 
can easily be overcome by a pre-diagonalization of hybridization terms.

\subsubsection{Flow-equation approach}
\label{App A1b}

Next, we reconsider  the  model \eqref{A.2} taking  advantage of the freedom discussed in Sec.~\ref{II.A} that the part 
$\mathbf{P}_{\lambda-\Delta\lambda} X_{\lambda,\Delta\lambda}$ of the generator with low energy transitions is not fixed 
in our diagonalization scheme. 
Therefore, instead of taking a vanishing $\mathbf{P}_{\lambda-\Delta\lambda} X_{\lambda,\Delta\lambda}$ as in the minimal transformation 
 let us use a non-vanishing $\mathbf{P}_{\lambda-\Delta\lambda} X_{\lambda,\Delta\lambda}$. 
Choosing a suitable expression 
we show below that the renormalization method in this case becomes identical  to  Wegner's continuous flow-equation 
method and  can be fully  understood in the framework of the present approach. 
 Thereby $\mathbf{P}_{\lambda-\Delta\lambda} X_{\lambda,\Delta\lambda}$ is chosen such that the part $\mathbf{Q}_{\lambda-\Delta\lambda} X_{\lambda,\Delta\lambda}$ can be neglected taking for $\mathbf{P}_{\lambda-\Delta\lambda} X_{\lambda,\Delta\lambda}$ 
an expression of  the same operator structure as Eq. \eqref{A.8},
\begin{equation}
\begin{split}
\label{A.16}
 {\bf P}_{\lambda -\Delta \lambda} X_{\lambda,\Delta\lambda} &=
  \sum_{{k}}
  A_{k}(\lambda,\Delta\lambda)  \,  \Theta_{k, \lambda} \Theta_{k, \lambda - \Delta\lambda} \\
&\times  \left(
    f_{k}^{\dagger} c_{k}^{} - 
    c_{k}^{\dagger} f_{k}^{}
  \right) ,
  \end{split}
\end{equation}
where only low-energy  excitations are considered. This is realized by the products of the two $\Theta$-functions and 
\begin{equation}
\label{A.17}
  A_{k}(\lambda,\Delta\lambda) = \Delta \lambda \,
   \alpha_{k}(\lambda,\Delta\lambda) \,
\end{equation} 
with 
\begin{equation}
\label{A.18}
\alpha_{k}(\lambda,\Delta\lambda) =  
  \frac{
    \left(
        \varepsilon_{{k},\lambda}^f  - 
        \varepsilon_{k,\lambda}^c
    \right)
    V_{k,\lambda}
  }{
    \kappa
    \Big[
      \lambda - \Big|
        \varepsilon_{{k},\lambda}^f - 
        \varepsilon_{k,\lambda}^c
      \Big|
    \Big]^{2} 
  }  .
\end{equation} 
The quantity $\kappa$ in the denominator is a free energy constant and was introduced  to 
ensure vanishing dimensionality of $A_{k}({\lambda, \Delta \lambda})$. Moreover, $A_{k}(\lambda,\Delta\lambda)$ 
is chosen proportional to $\Delta \lambda$ in order to reduce the impact of the actual value of 
$\Delta\lambda$ on the final results of the renormalization. 
Note that there  is no derivation of expression \eqref{A.18}. Instead, we have made use of the freedom to chose the low energy part of the generator arbitrarily. It turns out that Eq.~\eqref{A.18}  is indeed a reasonable choice. In particular, it will be shown that  in the limit of small $\Delta\lambda$  it leads to 
 a rapid but continuous  decay of the hybridization $V_{k,\lambda}$ and thus to a vanishing 
 'interaction' $\mathcal H_1$. The expected behavior with decreasing $\lambda$ is shown in Fig.~\ref{Fig_Fano}(b). Thus, the initial assumption that the part $\mathbf{Q}_{\lambda-\Delta\lambda} X_{\lambda,\Delta\lambda}$ can be neglected is justified by the particular choice \eqref{A.16} with the prefactors \eqref{A.17} and \eqref{A.18}.
 
In order to derive continuous renormalization equations, as is done in the flow equation method, we exploit the advantage that 
$A_{k}(\lambda, \Delta \lambda)$ is proportional to $\Delta \lambda$. Therefore, 
 one best uses  Eqs.~\eqref{A.10} and \eqref{A.11}, which are also valid in the present case, and find  in the limit $\Delta \lambda \rightarrow 0$
\begin{equation}
\label{A.19}
\frac{d\varepsilon_{{k},\lambda}^{(c,f)}}{d\lambda} = 
\pm 2 V_{{k},\lambda} \alpha_{k, \lambda}, 
\end{equation} 
where higher order terms in $V_{k,\lambda} \sim \Delta \lambda$ drop. 
A similar equation is also derived for $V_{{k}, \lambda}$,
\begin{equation}
  \label{A.21}
  \frac{d V_{{k},\lambda}}{d\lambda}
  = (\varepsilon_{{k},\lambda}^f - 
  \varepsilon_{{k},\lambda}^c)\, 
  \alpha_{{k}, \lambda}.
\end{equation} 
The renormalization equations \eqref{A.19} and \eqref{A.21} are solved by rewriting at first Eq.~\eqref{A.21}, 
\begin{eqnarray}
  \label{A.22}
  \alpha_{{k}, \lambda} &=& 
  \frac{1} {\varepsilon_{{k},\lambda}^f - 
  \varepsilon_{{k},\lambda}^c} 
  \frac{d V_{{k},\lambda}}{d\lambda} ,
\end{eqnarray} 
and by inserting this result into Eqs.~\eqref{A.19}. Using the property 
$
  \varepsilon_{{k}, \lambda}^f
  + \varepsilon_{{k}, \lambda}^c = \varepsilon_{{k}}^f
  + \varepsilon_{{k}}^c
$, which also follows 
from Eqs.~\eqref{A.19}, we obtain
\begin{eqnarray}
  \label{A.23}
  0 &=& \frac{d}{d\lambda} 
  \left\{
    (\varepsilon_{{k}, \lambda}^c)^2  -
    (\varepsilon_{{k}}^f +\varepsilon_{{k}}^c)
    \varepsilon_{{k}, \lambda}^c + V_{{k}, \lambda}^2
  \right\}.
\end{eqnarray} 
Eq.~\eqref{A.23} is easily integrated and leads for $\lambda \rightarrow 0$ 
to a quadratic equation for 
$\tilde{\varepsilon}_{{k}}^c = \lim_{\lambda \rightarrow 0}
\varepsilon_{{k}, \lambda}^c$. Its solution 
corresponds to the former result \eqref{A.14}, 
whereas $\tilde{\varepsilon}_{{k}}^f$ is found from 
$\varepsilon_{{k}, \lambda}^f
+ \varepsilon_{{k}, \lambda}^c = \varepsilon_{{k}}^f
+ \varepsilon_{{k}}^c$. \\

Finally let us study the $\lambda$-dependence of 
 $V_{{k}, \lambda}$. According to Eqs.~\eqref{A.22} and 
\eqref{A.23} $V_{{k}, \lambda}$ is governed by
\begin{eqnarray}
\label{A.24}
  \frac{d \ln V_{{k},\lambda}}{d\lambda}
  &=& \frac{(\varepsilon_{{k},\lambda}^f - 
  \varepsilon_{{k},\lambda}^c)^2}
  {\kappa [\lambda - |\varepsilon_{{k}, \lambda}^f -
  \varepsilon_{{k}, \lambda}^c|]^2   } \,
  \Theta(\lambda -|\varepsilon^f_{{k},\lambda }
  -\varepsilon^c_{{k},\lambda }| ). \nonumber \\
  &&
\end{eqnarray} 
One concludes:
\begin{enumerate}
  \item[(i)]
  The coupling matrix elements $V_{{k},\lambda}$ continuously decay 
  to smaller values when the cutoff energy $\lambda$ is lowered. Due to the denominator in Eq.~\eqref{A.24} the decay starts at $k$ values with the largest transition energy ($k=0$ and $k=2$ in Fig.~\ref{Fig_Fano}). In the $k$ range with the lowest transition energies around the intersection point of $\varepsilon_k$ and $\varepsilon_f$ ($k=1$ in Fig.~\ref{Fig_Fano}) the decay happens later but the renormalization of $\varepsilon_{{k},\lambda}^c$ and $\varepsilon_{{k},\lambda}^f$ is strongest (compare Fig.~\ref{Fig_Fano}). 
  \item[(ii)]
  For the particular $\lambda$ value, 
  $
    \lambda = 
    \big|
      \varepsilon_{{k}, \lambda}^f - 
      \varepsilon_{{k}, \lambda}^c
    \big|
  $,
  the denominator in Eq.~\eqref{A.24} guarantees that the  renormalized coupling strength $V_{{k},\lambda}$  reaches the value zero.  
  Thus, as claimed before, the hybridization completely vanishes. The continuous decay to zero  is different from the minimal transformation, since 
  thereby all excitations with non-zero energies are stepwisely integrated out during the renormalization procedure.  
\end{enumerate}    

\subsubsection{Comparison of the two approaches}
\label{App A1c}

In summary, we have shown that both renormalization schemes from the previous subsections lead to identical results, i.~e.~to the exact diagonal form for the  Hamiltonian. This was demonstated here for the example of a system without interaction. However, correlation and fluctuation effects can also be studied in a similar manner \cite{Kehrein_2006}.

The two schemes differ in their particular choices of the low energy part ${\mathbf P}_{\lambda - \Delta \lambda} X_{\lambda, \Delta \lambda}$ of the generator. While this part is set to zero within the so-called minimal transformation it can alternatively  be chosen such that  it is the {\it only remaining part} of the generator, so that the high-energy part ${\mathbf Q}_{\lambda - \Delta \lambda} X_{\lambda, \Delta \lambda}$ of the generator can be neglected. In this case the whole renormalization behavior is solely influenced by the low-energy  part. 
It also leads to the \textit{continuous} renormalization version  which is  equivalent to Wegner's flow equation method
and has the advantage that available computer subroutines can be used to solve the corresponding differential equations. 
In contrast,  the minimal transformation  is based on 
\textit{discrete} transformations leading to a system of coupled difference equations as Eqs.~\eqref{A.10} and \eqref{A.11}.
Moreover, it  has the advantage that the generator is completely fixed by the method itself whereas in the flow equation method
an appropriate choice for the generator must be made.

\section{Interacting many-particle systems}
\label{III}

Now we apply the developed concepts to  true many-particle systems with interactions. For this purpose  we employ two example systems which allow a mostly transparent presentation of specific  techniques which are necessary to treat interactions within our diagonalization scheme. Thereby, for the first example the concept of the minimal transformation and 
for the second example the flow equation method is used.

\subsection{Holstein model}
\label{III.A}

We start with the minimal transformation and consider this technique for the  spinless Holstein model (HM) in one dimension, which is perhaps the simplest realization of a strongly coupled
electron-phonon (EP) system. The Hamiltonian of the model describes  dispersionless 
longitudinal optical phonons which locally interact with electrons of  density
$n_i= c^\dag_i c^{}_i$ at site $i$ and reads 
\begin{equation}
\begin{split}
  \label{4.1}
  {\cal H} &=
  - t \sum_{\langle i,j\rangle} ( c_{i}^\dagger c_{j}^{} + \mathrm{h.c.} )
  + \omega_0 \sum_i  b_i^\dagger b_i^{}  \\
  &+  g \sum_i \; (b_i^\dagger + b_i^{})n_i .
\end{split}
\end{equation}
 Here, $c^{\dagger}_{i} (c_i^{})$ and $b^{\dagger}_{i} (b_i^{})$ denote the 
local creation (annihilation) operators of electrons and phonons.
The electron-phonon coupling constant and frequency of the Einstein mode are given by $g$ and $\omega_0$,
and $t$ is the electronic hopping constant.
 With increasing EP coupling $g$, the HM undergoes a quantum-phase transition from 
a metallic state to a charge-ordered insulating state. In particular, at half-filling the 
insulating state of the HM is a dimerized Peierls phase. 

The model is not exactly solvable. Therefore, a number of different analytical and
numerical methods have  been applied to the model: strong coupling expansions
\cite{HF_1983}, Monte Carlo simulations \cite{HF_1983, MHM_1996}, variational
\cite{ZFA_1989} and renormalization group \cite{HM_RG,BGL_1995} approaches, exact
diagonalization techniques \cite{HM_ED,FHW_2000}, density matrix renormalization
group \cite{BMH_1998,JZW_1999,FWH_2005} and dynamical mean-field theory \cite{MHB_2002}. 
However, most of these approaches are restricted in
their application. In particular, in numerical methods the infinite phononic Hilbert space (even for finite
systems) demands either the application of truncation schemes 
or involves reduction procedures. As will be shown below within the diagonalization method presented here the phononic 
Hilbert space is not reduced.

At first, we consider the same starting point as in the previous subsection and show that it leads to a reliable description of the metallic state in the Holstein model. In particular, according to Refs.~\cite{SHBWF_2005,SHB_2006_2}, such a treatment  allows  access to  the crossover  between the so-called  adiabatic and anti-adiabatic limit of the model. 

As introduced in Sec.~\ref{App A1} the method starts  with  an  {\it ansatz} for the renormalized Hamiltonian $\mathcal H_\lambda$ which has the particular  property that  the operator structure of the original Hamiltonian is kept and that only the parameters become  $\lambda$-dependent. 
In momentum space of the HM this {\it ansatz} $\mathcal{H}_{\lambda} = \mathcal{H}_{0,\lambda} + \mathcal{H}_{1,\lambda}$ has the following form, 
\begin{equation}
\begin{split}
  \label{4.2} 
\mathcal{H}_{0,\lambda} &=
  \sum_{k} \varepsilon_{k,\lambda} c^{\dagger}_{k}c^{}_{k} +
  \sum_{q} \omega_{q,\lambda} b^{\dagger}_{q} b^{}_{q} + E_{\lambda}, 
 \\
 \mathcal{H}_{1,\lambda} &=
  \frac{g}{\sqrt{N}}\sum_{k,q}
  \Theta_{k,q,\lambda} \,
  \left(
    b^{\dagger}_{q} c^{\dagger}_{k}c_{k+q}^{} +
    b_{q} c^{\dagger}_{k+q}c_{k}^{}
  \right) .
   \end{split}
\end{equation} 
Here  the $\Theta$-function $\Theta_{k,q,\lambda} =
\Theta( \lambda - |\omega_{q,\lambda} + \varepsilon_{k,\lambda} -
\varepsilon_{k+q,\lambda}|)$ in  $\mathcal H_{1,\lambda}$ guarantees  that only transitions 
with energies smaller than $\lambda$ contribute to the interaction at cutoff $\lambda$.  Within the concept of the minimal transformation the coupling coefficient $g$ is kept constant.
Moreover, Fourier transformed one-particle operators have been used for convenience. Note that the operator terms in $\mathcal{H}_{0,\lambda}$ resemble the ones in the respective term in Eq.~\eqref{A.7}. The main difference, however, lies in the coupling $\mathcal{H}_{1,\lambda}$ which here describes a real interaction instead of a simple hybridization.   

The renormalization equations are derived according to the concept introduced in Sec.~\ref{II.A}
by removing all transitions in a small energy shell between  $\lambda$ and  
a somewhat reduced cutoff $\lambda-\Delta\lambda$,   
$\mathcal{H}_{\lambda-\Delta\lambda} = 
  e^{X_{\lambda, \Delta\lambda}}\; \mathcal{H}_{\lambda}\;  
  e^{-X_{\lambda, \Delta\lambda}}. $
  For the generator 
  $X_{\lambda,\Delta\lambda}$ of the unitary transformation we use a similar {\it ansatz} as already considered for the exactly solvable model in Sec.~\ref{App A1}. For the HM it has the form 
\begin{equation}
\label{4.3}
  X_{\lambda,\Delta\lambda} = \frac{1}{\sqrt{N}}\sum_{k,q} A_{k,q}(\lambda,\Delta\lambda)
  \left(
    b^{\dagger}_{q} c^{\dagger}_{k}c^{}_{k+q} -
    b^{}_{q} c^{\dagger}_{k+q}c^{}_{k}
  \right) .
\end{equation}
However, in contrast to the situation in  Sec.~\ref{App A1} no exact analytical expression for the prefactor $A_{k,q}(\lambda,\Delta\lambda)$ can be found. Instead, we here determine these coefficients using the first order expression \eqref{2.13} of the generator. Thus, neglecting all higher orders the generator of the unitary transformation reads
$X_{\lambda,\Delta\lambda} =
  {\bf L}_{0,\lambda}^{-1} {\bf Q}_{\lambda -\Delta \lambda} \mathcal H_{1,\lambda}.$
  We will show below that this assumption  is justified as long as the width $\Delta\lambda$ is kept small compared to the starting value $\Lambda$ of the renormalization procedure. Using this formula  we find the following analytical expression for the coefficient $A_{k,q}(\lambda,\Delta\lambda)$ which is of first order with respect to the interaction parameter $g$,
\begin{equation}
\label{A_HM}
  A_{k,q}(\lambda,\Delta\lambda) =  
  \frac{g}{\omega_{q,\lambda} + \varepsilon_{k,\lambda} - \varepsilon_{k+q,\lambda}}
   \, \Theta_{k,q}(\lambda,\Delta \lambda). 
  \end{equation}
Here, $\Theta_{k,q}(\lambda,\Delta
  \lambda) = \Theta_{k,q,\lambda}\,\big (1 - \Theta_{k,q,\lambda
  - \Delta \lambda} \big)$ is the product of the two $\Theta$-functions 
which restrict allowed transitions to energies between $\lambda$ and $\lambda -\Delta \lambda$. 
Note that expression 
\eqref{4.3} corresponds to the minimal transformation, obeying
${\bf Q}_{\lambda -\Delta \lambda} X_{\lambda, \Delta \lambda} = X_{\lambda, \Delta \lambda}$.

The next step is to derive the renormalization equations using Eq.~\eqref{2.16}. Here we take advantage that in the perturbative expansion  for a 'small' renormalization step from
$\lambda$ to $\lambda - \Delta \lambda$ the higher orders can be neglected. The renormalization step is  considered as 'small' if in the actual  numerical evaluation 
the width $\Delta \lambda$ is chosen sufficiently small so that
 only a small  number of renormalization processes contribute
  within the interval $\Delta \lambda$. 
 Thus, roughly speaking, the  'smallness parameter' is  defined as 
 the relative coupling strength of the 'small' perturbation $\mathcal H_1$, however  multiplied by the  small 
 ratio of the number of  renormalization processes within $\Delta \lambda$ divided by their total number. 
 In this way, perturbation theory in Eq.~\eqref{2.16} should be well fulfilled.  
 
Evaluating the two second order commutator expressions in Eq.~\eqref{2.16},  
terms with four fermionic and bosonic one-particle operators show up which can not directly be attributed to the  operator terms of Eq.~\eqref{4.2}. Note that this difficulty did not appear in the hybridization model \eqref{A.2} since in that case  the coupling term ${\cal H}_1$ was quadratic. For the HM, and also for any other Hamiltonian with interactions, terms with more than two operators are usually generated. This problem is solved as follows. 
To restrict the renormalization scheme to terms already
included in {\it ansatz} \eqref{4.2}, a factorization approximation has to be
employed, 
\begin{align}
\label{4.5}
  c^{\dagger}_{k}c_{k} c^{\dagger}_{k-q}c_{k-q} &\approx
  c^{\dagger}_{k}c_{k} \langle c^{\dagger}_{k-q}c_{k-q} \rangle +
  \langle c^{\dagger}_{k}c_{k} \rangle c^{\dagger}_{k-q}c_{k-q} \nonumber \\
  &-\langle c^{\dagger}_{k}c_{k} \rangle
  \langle c^{\dagger}_{k-q}c_{k-q} \rangle, \\[1ex]
  b^{\dagger}_{q} b_{q} c^{\dagger}_{k}c_{k} &\approx
  b^{\dagger}_{q} b_{q} \langle c^{\dagger}_{k}c_{k} \rangle +
  \langle b^{\dagger}_{q} b_{q} \rangle c^{\dagger}_{k}c_{k}  -
  \langle b^{\dagger}_{q} b_{q} \rangle
  \langle c^{\dagger}_{k}c_{k} \rangle , \nonumber 
\end{align}
which means that operators are partially replaced by  expectation values. This step allows to  trace back the
new operator terms to operator expressions which are already present  in $\mathcal H_\lambda$. 
Hence, the resulting renormalization equations will  contain expectation values which have to be
calculated separately. In principle, these expectation values should be defined with
respect to $\mathcal{H}_{\lambda}$, because the factorization 
is done for each renormalization step between $\lambda$ 
and $\lambda -\Delta \lambda$. 
However,  $\mathcal{H}_{\lambda}$ may
still contain interaction terms which prevent a straightforward  evaluation of 
expectation values. The easiest way to circumvent this difficulty would be to
neglect all interaction terms in $\mathcal{H}_{\lambda}$ and to use instead 
the diagonal unperturbed part $\mathcal{H}_{0,\lambda}$. This approach has been 
applied for instance in \cite{SHBWF_2005}, where single-particle excitations and
phonon softening of the Holstein model were studied. However, often 
interactions are crucial. For this reason,
it has turned out that expectation values should best be defined with  
the full Hamiltonian $\mathcal{H}$ which includes the full interaction $\mathcal H_1$.
As is described in the following they have to be 
determined self-consistently together with the renormalization equations of  
$\mathcal H_\lambda$. 

The renormalization equations for
$\varepsilon_{k,\lambda}$, $\omega_{q,\lambda}$, and
$E_{\lambda}$ are found by comparing the resulting expression
for $\mathcal H_{\lambda -\Delta \lambda}$  
with {\it ansatz} \eqref{4.2}, where  $\lambda$ is 
replaced by $\lambda - \Delta \lambda$. One finds \cite{SHBWF_2005},    
\begin{align}
  \label{4.6}
    &\varepsilon_{k,\lambda - \Delta \lambda} - \varepsilon_{k,\lambda} \nonumber \\
  &= \frac{1}{N} \sum_{q} \left( n_q^{\rm b} + n_{k+q}^{\rm c} \right) 
  \frac{g^2 \Theta_{k,q}(\lambda,\Delta \lambda)}
  {\omega_{q,\lambda} + \varepsilon_{k,\lambda} - \varepsilon_{k+q,\lambda}} \nonumber
  \\
  &- \frac{1}{N} 
  \sum_{q} \left( n_q^{\rm b} - n_{k-q}^{\rm c} + 1\right) 
  \frac{g^2 \Theta_{k-q,q}(\lambda,\Delta \lambda)}
  {\omega_{q,\lambda} + \varepsilon_{k-q,\lambda} - \varepsilon_{k,\lambda}},
\end{align}
and
\begin{align}
  \label{4.7}
   & \omega_{q,\lambda - \Delta \lambda} - \omega_{q,\lambda} \nonumber \\
    &=
  \frac{1}{N} \sum_{k} \left( n_k^{\rm c} 
   - n_{k+q}^{\rm c} \right) \frac{g^2 \Theta_{k,q}(\lambda,\Delta
  \lambda)}{\omega_{q,\lambda} + \varepsilon_{k,\lambda} 
  - \varepsilon_{k+q,\lambda}}  ,
\end{align}
and a similar equation for $E_\lambda$.
The renormalization equations \eqref{4.6} and \eqref{4.7}  depend on the yet unknown 
expectation values $n_k^{\rm c} =\langle c_{k}^{\dagger} c_{k} \rangle$ and
$n_q^{\rm b} = \langle b_{q}^{\dagger} b_{q} \rangle$ which arise from the 
factorization approximation \eqref{4.5}. As discussed above  they are best evaluated with
respect to the full Hamiltonian $\mathcal{H}$. Using Eq.~\eqref{2.21},
i.~e.~$\langle\mathcal{A}\rangle =  \lim_{\lambda\rightarrow 0}\langle
   \mathcal{A}_{\lambda}
 \rangle_{\mathcal{H}_{\lambda}} $,
where $ \mathcal{A}_{\lambda} = e^{X_\lambda} \mathcal A e^{-X_\lambda}$, 
an additional set of renormalization equations for $\mathcal A_\lambda$ should be derived:
Exploiting
$\langle (c_{k}^\dag c_{ k})_\lambda\rangle_{\mathcal H_\lambda} = 
\langle c_{ k, \lambda}^\dag c_{ k, \lambda}\rangle_{\mathcal H_\lambda}$ and
$\langle (b_{q}^\dag b_{ q})_\lambda\rangle_{\mathcal H_\lambda} = 
\langle b_{ q, \lambda}^\dag b_{ q, \lambda}\rangle_{\mathcal H_\lambda}$
we start from the following {\it ansatz} for the $\lambda$-dependent
fermionic and bosonic one-particle operators, 
 \begin{align}
 \label{4.8}
  c_{k,\lambda}^{\dagger} &=
  \alpha_{k,\lambda} \, c_{k}^{\dagger} + 
  \sum_{q} \left(
    \beta_{k,q,\lambda} \, c_{k+q}^{\dagger} b^{}_{q} + 
    \gamma_{k,q,\lambda} \, c_{k-q}^{\dagger} b_{q}^{\dagger}
  \right) , \\
 \label{4.9}
  b_{q,\lambda}^{\dagger} &=
  \phi_{q,\lambda} \, b_{q}^{\dagger} + \eta_{q,\lambda} \, b^{}_{-q} +
  \sum_{k} \psi_{k,q,\lambda} \, c_{k+q}^{\dagger} c^{}_{k} . 
\end{align}
Here the operator structure is suggested by the low order expansion 
of $c^\dag_{k, \lambda}$ and $b^\dag_{q, \lambda}$ in terms of 
$X_{\lambda, \Delta \lambda}$  \cite{SHBWF_2005}. 
The renormalization equations for the $\lambda$-dependent parameters in Eqs.~\eqref{4.8} and \eqref{4.9}
are found from relation \eqref{2.23}. For instance, the equations for the parameters 
$\phi_{q,\lambda}$, $\eta_{q,\lambda}$, and $\psi_{k,q,\lambda}$ of the phonon operator 
$b^\dag_{q, \lambda}$ read
\begin{equation}
\begin{split}
\label{4.10}
 \phi_{q, \lambda -\Delta\lambda} - \phi_{q,\lambda} &=  \sum_k \Big[
n_{k,q} A_{k,q}(\lambda, \Delta \lambda)\psi_{k,q,\lambda} \\
&- \frac{1}{2} n_{k,q} A^2_{k,q}(\lambda, \Delta \lambda) \phi_{q,\lambda} 
\Big],\
\end{split}
\end{equation}
\begin{equation}
\begin{split}
\label{4.11}
 \eta_{q, \lambda -\Delta\lambda} - \eta_{q,\lambda} &=  - \sum_k \Big[
n_{k,q} A_{k,q}(\lambda, \Delta \lambda)\psi_{k,q,\lambda}   \\
&+ \frac{1}{2} n_{k,q} A^2_{k,q}(\lambda, \Delta \lambda) \eta_{q,\lambda} 
\Big],
\end{split}
\end{equation}
and
\begin{equation}
\label{4.12}
\psi_{k,q,\lambda- \Delta \lambda} - \psi_{k,q,\lambda} = 
-\frac{1}{\sqrt N} A_{k,q}(\lambda, \Delta \lambda)\big( \phi_{k,q} + \eta_{k,q} \big) ,
\end{equation}
where we have defined $n_{k,q} = \langle c_k^\dag c_k\rangle -  \langle c_{k+q}^\dag c_{k+q}\rangle$. Similar renormalization equations are found for the parameters of $c^\dag_{k,\lambda}$.

The renormalization equations \eqref{4.10}-\eqref{4.12} for the parameters of $b_{q, \lambda}^\dag$  and 
$c_{k,\lambda}^\dag$   have to be solved self-consistently together with equations \eqref{4.6} and \eqref{4.7} for the parameters  of $\mathcal H_\lambda$, subject to the respective initial conditions 
(at cutoff $\lambda = \Lambda$),
\begin{align}
\label{4.13}
\varepsilon_{k, \Lambda} &= \varepsilon_{k} , & \omega_{q,\Lambda} &= \omega_0 , & E_\Lambda &= 0 , \nonumber \\
\alpha_{k, \Lambda} &= 1 ,  & \beta_{k,q, \Lambda} &= 0 ,& \gamma_{k,q, \Lambda} &= 0 ,\nonumber \\
\phi_{q, \Lambda} &= 1  , & \eta_{q, \Lambda} &= 0 , & \psi_{k,q, \Lambda} &= 0 \nonumber
\end{align}
($\varepsilon_k= -2t \cos{k} -\mu$). The numerical evaluation of the coupled renormalization equations 
starts from some chosen values for the expectation values. With 
this choice the evaluation cycle begins at cutoff $\Lambda$ 
and proceeds step by step until $\lambda = 0$ is reached. 
The limit $\lambda = 0$ allows to re-calculate all
expectation values, and the renormalization procedure starts again
with the improved expectation values 
by reducing again the cutoff from $\Lambda$ to $\lambda=0$. 
After a sufficiently large number of such cycles, the expectation values are
converged and the renormalization equations have been solved
self-consistently. As the final result  we obtain an effectively free model,
 $\tilde{\mathcal{H}} =
  \sum_{k} \tilde{\varepsilon}_{k}
  c^{\dagger}_{k}c^{}_{k} +
  \sum_{q} \tilde{\omega}_{q}
  b^{\dagger}_{q} b^{}_{q} +
  \tilde{E}$,
where we have again used tilde symbols for the fully renormalized quantities, 
$\tilde{\varepsilon}_{k}=\lim_{\lambda\rightarrow 0}\varepsilon_{k,\lambda}$,
$\tilde{\omega}_{q}=\lim_{\lambda\rightarrow 0}\omega_{q,\lambda}$, and 
$\tilde{E}=\lim_{\lambda\rightarrow 0}E_{\lambda}$. Analogous expressions 
are also found for the fully  renormalized quantities $c^\dag_{k,\lambda=0}$ and $b^\dag_{q, \lambda=0}$. 
Note that the fully renormalized Hamiltonian is diagonal so that any expectation value with 
$\tilde{\mathcal H}$ can be evaluated.  

A trivial counterexample, where 
the present approach may fail is the case of a flat energy dispersion of $\mathcal H_0$. For instance,
 for a system  $\mathcal H_0$ with discrete eigenvalues $E_n$ this might lead to a 
 large amount of renormalization processes in some small intervals $\Delta \lambda$, so that 
 the present renormalization treatment has to be modified. 
  Apart from the last case, the discussed method has turned out to give excellent results 
for quite a number of many-particle problems, which are valid  far beyond the range of validity of 
usual perturbation theory.

At first, let us show for the example of the HM that a numerical evaluation of the renormalization equations can provide a comprehensive understanding of the relevant  physical processes in  interacting many-particle systems with relatively large system size near the thermodynamic limit. We particularly show that reasonable agreement with other numerical techniques is obtained. For the HM the results of the numerical evaluation will be discussed for half-filling. Thereby three different  
cases have to be distinguished: (i) The adiabatic case $\omega_{0} \ll t$, 
(ii) the intermediate case $\omega_0 \approx t$ 
and (iii) the anti-adiabatic case  $\omega_{0} \gg t$.   

\begin{figure*}[t]
\centering           
\includegraphics[width=\textwidth]{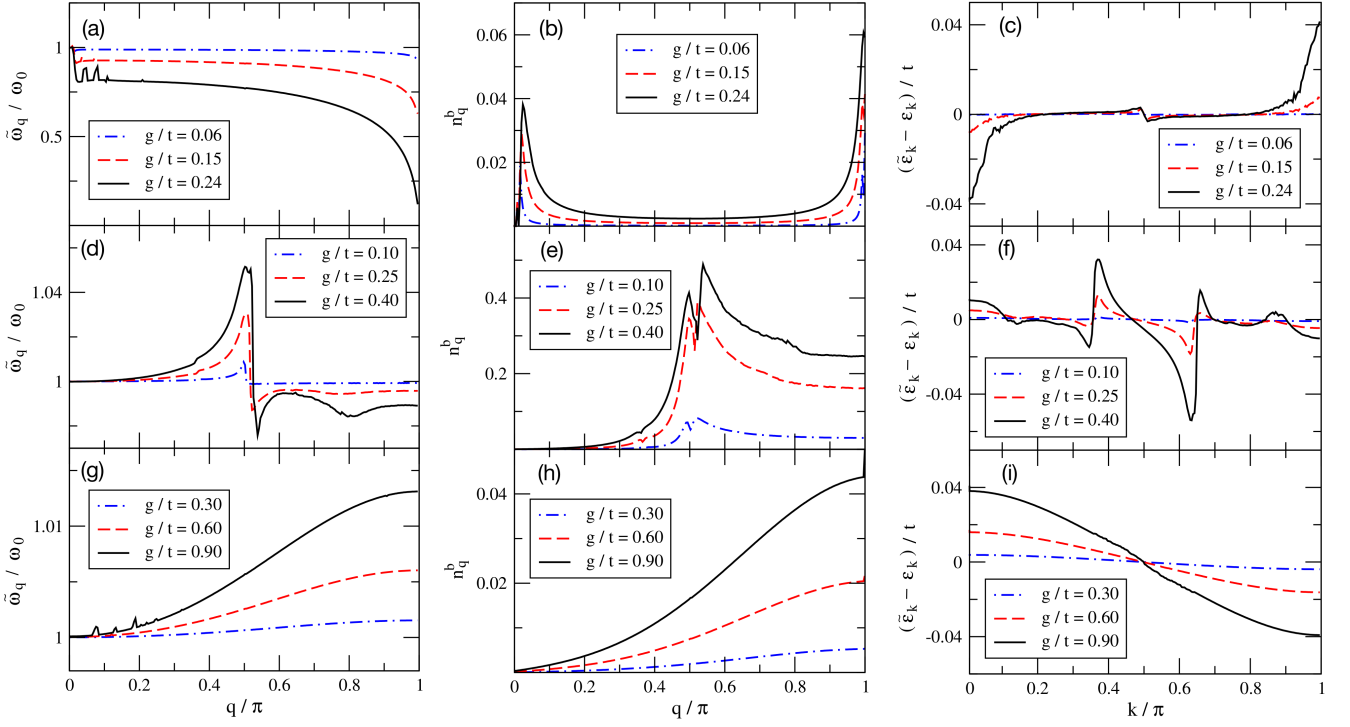}
  \caption{
     Results obtained from the numerical evaluation of the renormalization equations for the spinless Holstein model in the adiabatic (a-c), intermediate (d-f), and anti-adiabatic case (g-i).
    {\it Left:} Bosonic quasi-particle energies $\tilde{\omega}_{q} /
    \omega_0$ at half-filling as a function of $q$ for 
    different values of the EP coupling $g$. {\it Middle:} Phonon distribution $n_q^{\rm b} =
    \langle b_q^\dag b_q \rangle$ as a function of $q$ for the same parameters
    as in the left panel. {\it Right:} Fermionic quasi-particle energies $(\tilde{\varepsilon}_{k}
    - \varepsilon_{k}) / t$ as a function of $k$ for the same parameters
    as in the left and middle panel. Here $\varepsilon_{k}$ is the
    original electronic dispersion. For the phonon energy the following parameter values were used:  $\omega_0/t= 0.05$ in the adiabatic case, $\omega_0/t = 2.8$ in the intermediate case, $\omega_0/t = 6.0$ in the anti-adiabatic case.
  }
  \label{Fig_phonon_dispersion_2}
\end{figure*}

{\it Adiabatic case:\ }
The results for the adiabatic case are shown in Fig.~\ref{Fig_phonon_dispersion_2}(a-c).
In  Fig.~\ref{Fig_phonon_dispersion_2}(a) the phononic
quasi-particle energy $\tilde{\omega}_{q}$ shows a weakening due to a
gain in dispersion for increasing coupling between electronic and 
phononic degrees of freedom, in particular around $q=\pi$. If the 
coupling exceeds a critical value $g_{c}$ non-physical negative
energies at $q=\pi$ occur, signaling the break-down of the
present description for the metallic phase at the quantum-phase transition to
the insulating Peierls state. 
%
Note that the vanishing of the phonon mode also allows to determine   
the critical EP coupling $g_{c}$ of the phase transition 
(see Ref.~\cite{SHBWF_2005}). For example, at half-filling and $\omega_0 = 0.1t$, 
a value of $g_{c}=0.31t$ is found, which is somewhat
larger than that found by DMRG in  
Refs.~\cite{BMH_1998} and \cite{FWH_2005}, which is $g_{c}=0.28t$.

Fig.~\ref{Fig_phonon_dispersion_2}(b) shows the phonon distribution
$n_q^{\rm b} = \langle b_q^\dag b_q \rangle$ for the same parameter values
as in Fig.~\ref{Fig_phonon_dispersion_2}(a). There are two pronounced
maxima found at wave numbers $q = \pi$ and $q \approx 0$. The peak at $q
= \pi$ is directly connected to the softening of $\tilde{\omega}_q$ at 
the zone boundary and can be considered as
a precursor  of the transition to a dimerized state. For 
the exact critical EP coupling $g= g_c$ $(\approx 0.24 \  \mbox{for \ } \omega_0/ t = 0.05 )$  a divergency of
$n_q^{\rm b}$ appears at $q = \pi$. 
The second peak around $q\approx 0$ follows from renormalization
contributions which also become strong for small $q$. 
This will be explained in more detail below.

Finally, in Fig.~\ref{Fig_phonon_dispersion_2}(c) the renormalized fermionic
one-particle energy $\tilde{\varepsilon}_k$ is shown in relation to  
the original dispersion $\varepsilon_k$ for the same parameter
values as in Fig.~\ref{Fig_phonon_dispersion_2}(a). 
Though the absolute changes are quite small, the difference between
$\tilde{\varepsilon}_k$ and $\varepsilon_k$ is strongest in the
vicinity of $k=0$ and $k=\pi$. In particular, we find $\tilde{\varepsilon}_k <
\varepsilon_k$ for $k=0$ and $\tilde{\varepsilon}_k >
\varepsilon_k$ for $k=\pi$, so that the renormalized
bandwidth becomes somewhat larger than $4t$, which is the original bandwidth. 

{\it Intermediate case:}
The results for the intermediate case ($\omega_0 / t = 2.8$) are shown in Figs.~\ref{Fig_phonon_dispersion_2}(d-f). In contrast to the
adiabatic case, the renormalized phonon energy   $\tilde{\omega}_q$
in Fig.~\ref{Fig_phonon_dispersion_2}(d) 
has a noticeable 'kink' at some intermediate wave vector $q_k$, which is a   
specific feature of the intermediate case. Thereby $q_k$ strongly
depends on the initial phonon energy $\omega_0$.
It is characterized by a strong
renormalization of the phonon energy in a small $q$-range around $q_k$, where 
$\tilde{\omega}_q / \omega_0 > 1$  for $q < q_k$ and $\tilde{\omega}_q /
\omega_0 < 1$ for $q > q_k$ holds. The origin of this
feature will be discussed below.    

Similar to $\tilde{\omega}_q$, also the phonon distribution $n_q^{\rm b}$ in 
Fig.~\ref{Fig_phonon_dispersion_2}(e) shows a pronounced structure of considerable
weight around $q_k$. Finally, in 
Fig.~\ref{Fig_phonon_dispersion_2}(f), where  the difference of the  fermionic
one-particle energies $(\tilde{\varepsilon}_k - \varepsilon_k)$ is shown, 
again a remarkable structure is found, though the absolute changes are  quite
small for the present $g$-values. 
  

{\it Anti-adiabatic case:}
Finally, we discuss the results for the anti-adiabatic case $\omega_0
\gg t$. In Figs.~\ref{Fig_phonon_dispersion_2}(g-i) a value of
$\omega_0 / t = 6.0$ is used. As most important feature 
a stiffening of the renormalized
phonon frequency  $\tilde{\omega}_q$ is found in Fig.~\ref{Fig_phonon_dispersion_2}(g)
instead of a softening as in the adiabatic case.
In particular, at $q=\pi$ no softening
of the phonon modes occurs.
Moreover, no large renormalization contributions occur 
in any limited $q$-space regime, leading to peak-like structures.  
Instead an overall smooth behavior is found in the entire
Brillouin zone. Note that  also  the phonon distribution $n_q^{\rm b}$ in Fig.~\ref{Fig_phonon_dispersion_2}(h)
shows a smooth increase with a maximum at $q=\pi$. The lack of strong 
peak-like structures in $q$ space indicates that there is 
no phonon mode which gives rise to dominant contributions 
to the renormalization processes.  

If one compares the renormalized electronic bandwidth for the anti-adiabatic
case [Fig.~\ref{Fig_phonon_dispersion_2}(i)]
with that of the adiabatic case [Fig.~\ref{Fig_phonon_dispersion_2}(c)], 
one observes a reduction of the bandwidth. 
This indicates the
tendency to localization in the anti-adiabatic case. It also indicates that the
metal-insulator transition in this limit can be understood as
the formation of small immobile polarons with electrons surrounded by
clouds of phonon excitations. A renormalized 
one-particle excitation like the quantity $\tilde{\varepsilon}_k$ 
corresponds to a quasi-particle of the coupled many-particle system. 
A completely flat  
$\tilde{\varepsilon}_k$ should be found 
in the insulating charge density wave regime for still larger $g$. 

It might be worth to mention some extensions of the presented treatment in electron-boson systems which were studied in the past in the context of the projector-based renormalization method (PRM). In the preceding subsection we have considered the Holstein model as a particular  realization of coupled electron-phonon systems and have focused our attention to the  metallic  state. Moreover, in Ref.~\cite{SHB_2006_1} we have studied the quantum phase transition of the $1d$ Holstein model from the metallic to an insulating charge-ordered
phase. In this study a unified concept that covers both the metallic and the insulating phase  in the adiabatic limit has been developed.  
In two dimensions the electron-phonon interaction may additionally  lead to the formation of Cooper pairs giving rise to BCS superconductivity.  In Ref.~\cite{HB_2003} a microscopic derivation of the BCS-gap equation could be achieved using the technique described above. 
The quantum phase transition between superconductivity and charge order in the two-dimensional half-filled Holstein model is a further example which was addressed by the present formalism.   
In Ref.~\cite{SHB_2009} it was shown how such a competition of two ordering phenomena can be treated within the PRM framework. Thereby a crossover behavior between a purely superconducting state and a charge density wave was found, including a well-defined parameter range 
where superconductivity and lattice distortion coexists. 

Note  that the developed technique can also be applied without much additional effort to a generalized fermion-boson system. More specifically, we have investigated in the past a general fermion-boson interaction within the Edwards model~\cite{Ed06} which was originally proposed  as an elementary but non-trivial  fermion-boson model with the aim 
to describe quantum transport. Although the captured physics of the Edwards model is in several aspects different from that of the Holstein model the application of the PRM to solve both model Hamiltonians turned out to be conceptually similar and reliable. In particular, for the metallic state away from half-filling and dimension $d=1$ it was found 
that this model shows 
electronic phase separation for a certain parameter range (compare Ref.~\cite{SBF10} and the subsequent discussion in Ref.~\cite{ESBF12}). Moreover, for $d=2$ and half-filling 
a competition between unconventional superconducting pairing and charge density wave formation was found (Ref.~\cite{Cho2016}).

\subsection{Extended Falicov-Kimball model} 
\label{III.B}

In the previous subsection we have demonstrated that the diagonalization scheme based on the minimal transformation is able to solve models where fermions are coupled to a system of bosons, thereby generating  an effective interaction between fermions mediated by bosons and vice versa. Now we show that the generalized diagonalization method can also be used to study models where an explicit fermion-fermion interaction is given from the beginning and not necessarily provided by other degrees of freedom. 
As an example we consider the extended Falicov-Kimball model (EFKM) for two kinds of electrons as introduced in Ref.~\cite{Batista_2002}. The corresponding Hamiltonian is an extension of the original Falicov-Kimball model \cite{Falicov_1969} with finite dispersion 
for both types of the electron species. It has been shown in \cite{Batista_2002,Batista_2004} that such an extension leads to a novel ferroelectric state  in the strong-coupling and mixed-valence regime. In particular, the anticipated "BCS-Bose-Einstein condensate (BEC) crossover" scenario, 
connecting the physics  of BCS superconductivity with that of BEC's,  is of vital importance. The EFKM can capture this physics since it includes a direct $f$-$f$
hopping term \cite{Batista_2002} which provides at the same time  a  more realistic description than the entirely localized $f$ electrons in the conventional Falicov-Kimball model which  has already been studied within the PRM in Ref.~\cite{Becker2007}.    
 
Below we derive the basic formalism to integrate out the fermion-fermion interaction of the EFKM within our method. Here we use the non-zero low energy part ${\bf P}_{\lambda-\Delta \lambda} X_{\lambda, \Delta \lambda} $ of the generator of our unitary transformation \eqref{2.8} as discussed in Subsecs.~\ref{II.A.2} and \ref{App A1b}. Allowing for non-zero contributions according to Fig.~\ref{Fig_matrix_el} the interaction term is integrated out continuously and not in discrete steps as before.  
In this way the theoretical treatment leads to reliable results for photoemission spectra of the EFKM which can be used to probe the signatures of the excitonic condensate. The basic concepts of the theoretical approach in this subsection and 
in the main conclusions are taken over from Ref.~\cite{PBF2010}. For a simplified treatment we consider here the one dimensional case only, however the formalism is  also valid in higher dimensions \cite{PBF2010}.

The Hamiltonian for the EFKM in one dimension is written
\begin{equation}
\label{5.1}
\mathcal{H}=\sum_{k}\bar{\varepsilon}^c_{k}c^\dagger_{k}c^{}_{k}
+\sum_{k}\bar{\varepsilon}^f_{k}f^\dagger_{k}f^{}_{k}
+\sum_{i}Un^c_in^f_i,
\end{equation}
where   
$c^\dagger_{k}$ ($c^{}_{k}$) and 
$f^\dagger_{k}$ ($f^{}_{k}$)
are the creation (annihilation) operators in momentum ($k$-) 
space of spinless  $c$ and $f$-electrons, respectively,
and $n^c_i$ and $n^f_i$ are the corresponding occupation number operators
in real space. The  $c(f)$-fermion dispersion is
 \begin{equation}
 \label{5.2}
\bar{\varepsilon}^{c(f)}_{k}=\varepsilon_{}^{c(f)}-t^{c(f)}_{}
\gamma^{}_{k}-\mu
\end{equation}
with on-site energy $\varepsilon^{c(f)}_{}$ and chemical potential $\mu$. In the tight-binding limit,  we have 
$\gamma^{}_{ k}=2  \cos k$. 
The sign of $t^ct^f$  determines whether we deal with 
a direct ($t^ct^f <0$) or indirect ($t^ct^f>0$) band gap situation. 
Usually, the $c$-electrons are considered to be `light' and
their hopping integral is taken to be the unit of energy ($t^c=1$), 
while the $f$-electrons are `heavy', i.~e., $|t^f|< 1$. 
For $t^f\equiv 0$ (dispersionless $f$ band), 
the local $f$-electron number is strictly conserved~\cite{SC08}.     
The third term in Hamiltonian~\eqref{5.1} represents the 
Coulomb interaction between $c$ and $f$ electrons at the same 
lattice site. Hence, if the $c$ and $f$ bands are degenerate, 
$\varepsilon^{c}=\varepsilon^f$ and $t^c=t^f$, 
the EFKM reduces to the standard Hubbard model. 

We look for a non-vanishing excitonic expectation value 
$\langle c^\dagger f^{}\rangle$, indicating a kind of 
spontaneous symmetry breaking due to the pairing of $c$ 
electrons ($t^c>0$) with $f$ holes ($t^f <0$). We introduce two-particle interaction operators 
in momentum space,
$a^{}_{k_1k_2k_3}=c^\dagger_{k_1}
c^{}_{k_2}
f^\dagger_{k_3}f^{}_{k_1+k_3-k_2}$,
and rewrite the EFKM Hamiltonian~\eqref{5.1} in a normal-ordered 
form~\cite{Kehrein_2006},
\begin{equation}
\label{5.4}
\begin{split}
\mathcal{H} &= \sum_{k}\varepsilon^c_{k}:c^\dagger_{k}c^{}_{k}:
+\sum_{k}\varepsilon^f_{k}:f^\dagger_{k}f^{}_{k}:\\
&- \sum_{k}\left(\Delta :f^\dagger_{k}c^{}_{k}:+\, \textrm{h.c.}\right)
+\frac{U}{N}\sum_{k_1k_2k_3}
:a^{}_{k_1k_2k_3}:,
\end{split}
\end{equation}
where 
\begin{equation}
 \Delta  = \frac{U}{N}\sum_{k}{d}^{}_{k}\quad \mbox{with} \quad  
 {d}^{}_{k}=\langle c^\dagger_{k}f^{}_{k}\rangle .
\label{5.5}
\end{equation}
In the normal-ordered representation from operators $\mathcal A$ all possible factorizations are subtracted,
for instance  $ :~c_{k}^\dag c_{k}: = c_{k}^\dag c_{k} -\langle c_{k}^\dag c_{k}  \rangle$. 
Below, the quantity ${d}^{}_{k}$ plays the role of an order parameter. 
 Allowing a non-zero $d_{k}$, the symmetry 
of the Hamiltonian is explicitly broken, and iterating  
the self-consistency equation derived below will readily 
give  (meta-) stable solutions~\cite{KW06}. 
In Hamiltonian~\eqref{5.4}, the on-site energies were  
shifted by a Hartree term,
\begin{equation}
  \varepsilon^{c(f)}_{k}= \bar{\varepsilon}^{c(f)}_{k}
+U\langle n^{f(c)}\rangle,
\label{5.6}
\end{equation}
where $ n^c =\frac{1}{N}\sum_{k} 
\langle c^\dagger_{k} c^{}_{k}\rangle$,  $ n^f =\frac{1}{N}\sum_{k} 
\langle f^\dagger_{k} f^{}_{k}\rangle$
are the particle number densities of $c$ and $f$ electrons for  
a system  with $N$ lattice sites. In what follows, 
we consider the half-filled band case, i.~e., we fix the 
total electron density to $n= n^c +  n^f =1$.

The decomposition of the original Hamiltonian, $\mathcal{H}=\mathcal{H}_0+\mathcal{H}_1$, according to Eq.~\eqref{2.1} is here chosen in the form
\begin{equation*}
\begin{split}
\mathcal{H}_0&= \sum_{k}\varepsilon^c_{k}:c^\dagger_{k}c^{}_{k}:
+\sum_{k}\varepsilon^f_{k}:f^\dagger_{k}f^{}_{k}: \\
&+ \sum_{k}\left(\Delta :f^\dagger_{k}c^{}_{k}:+\, \textrm{h.c.}\right),
\end{split}
\end{equation*}
and
\begin{equation*}
\mathcal{H}_1=\frac{U}{N}\sum_{k_1k_2k_3}
:a^{}_{k_1k_2k_3}:.
\end{equation*}
Note that the hybridization term $\propto \Delta$
is included in ${\mathcal H}_0$ since it can exactly be taken into account by diagonalization of the fermion basis (compare Sec.~\ref{App A1}). Instead the perturbation ${\mathcal H}_1$ now only contains the fluctuating operator 
part of the Coulomb repulsion $\propto U$. 

Following the ideas of Sec.~\ref{II.A}, we decompose the renormalized Hamiltonian 
${\mathcal H}_\lambda$, after all transitions with energies larger than $\lambda$
are integrated out, into
$\mathcal{H_\lambda}=\mathcal{H}_{0,\lambda}+\mathcal{H}_{1,\lambda}$
with
\begin{align}
\label{5.11}
\mathcal{H}_{0,\lambda}&=\sum_{k}\varepsilon^c_{k,\lambda}:c^\dagger_{k}c^{}_{k}:   
+\sum_{k}\varepsilon^f_{k,\lambda}:f^\dagger_{k}f^{}_{k}:+E_\lambda \nonumber \\
&+\sum_{k}\left(\Delta^{}_{k,\lambda}:f^\dagger_{k}c^{}_{k}:+\,\textrm{h.c.}\right), \\
\label{5.12}
\mathcal{H}_{1,\lambda}&=\frac{1}{N}\mathbf{P}_\lambda \sum_{k_1k_2k_3}
U^{}_{k_1k_2k_3,\lambda}\,:a^{}_{k_1k_2k_3}:.
\end{align}
Here, $\mathbf{P}_\lambda$ again projects on all low-energy transitions with respect to the 
unperturbed Hamiltonian $\mathcal{H}_{0,\lambda}$ which are smaller than $\lambda$. 
Due to renormalization 
all prefactors in Eqs.~(\ref{5.11}), \eqref{5.12}  may now depend on the momentum  $k$ and on the 
energy cutoff $\lambda$.  
The quantity $E_\lambda$ is again an energy shift which enters during the renormalization
procedure. 
In order to evaluate the action of the superoperator $\mathbf{P}_\lambda$ 
on the interaction operator in $\mathcal{H}_{1,\lambda}$, in principle
one has to decompose the fluctuation operators $:a^{}_{k_1k_2k_3}:$
into eigenmodes of ${\mathcal H}_{0,\lambda}$, which would require a prior diagonalization of 
${\mathcal H}_{0,\lambda}$. 
However, here we consider only values of $U$ for which the mixing parameter $\Delta_{{k},\lambda}$ in Eq.~\eqref{5.11} 
is small compared to the energy difference  
  $|\varepsilon_{{k}, \lambda}^c - \varepsilon_{{k}, \lambda}^f|$. 
  This follows from the Hartree shifts of the one-particle energies in Eq,~\eqref{5.6}. 
  Thus, using as approximation
 ${\mathbf L}_{0,\lambda} c_{k}^\dagger = \varepsilon_{k}^c  c_{k}^\dagger$  
 and 
 ${\mathbf L}_{0,\lambda} f_{k}^\dagger = \varepsilon_{k}^f  f_{k}^\dagger$,
 we conclude 
 \begin{equation}
\label{5.13}
\mathcal{H}_{1,\lambda}=\frac{1}{N}\sum_{k_1k_2k_3}\,
\Theta(\lambda-|{\omega}^{}_{{k}_1{k}_2k_3,\lambda}|)\, 
U^{}_{k_1k_2k_3,\lambda}  :a^{}_{k_1k_2k_3}:\,,
\end{equation}
where 
${\omega}^{}_{{k}_1{k}_2{k}_3,\lambda}=
\varepsilon^c_{{k}_1,\lambda}-\varepsilon^c_{{k}_2,\lambda}
+\varepsilon^f_{{k}_3,\lambda}-\varepsilon^f_{{k}_1+{k}_3-{k}_2,\lambda}$
is the approximate excitation energy of $:a^{}_{k_1k_2k_3}:$, i.~e.
\begin{eqnarray}
\label{5.15}
{\mathbf L}_{0, \lambda}\, :a^{}_{k_1k_2k_3}:  = 
{\omega}^{}_{{k}_1{k}_2{k}_3,\lambda} 
 :a^{}_{k_1k_2k_3}:.
\end{eqnarray}
The $\Theta$-function in Eq.~\eqref{5.13} ensures that  
only transitions with excitation energies smaller than $\lambda$
remain in ${\mathcal H}_{1,\lambda}$.

By integrating out all transitions between the cutoff $\Lambda$ of the original model and $\lambda=0$,
all $\lambda$-dependent parameters of the original model will become fully renormalized. 
To find their $\lambda$-dependence,
we derive renormalization equations for the parameters $\varepsilon_{k,\lambda}^c$, $\varepsilon_{k, \lambda}^f$, $\Delta_{{k}, \lambda}$, and $U^{}_{k_1k_2k_3,\lambda}$. 
The initial parameter values are determined by the original model ($\lambda= \Lambda$),
\begin{align}
\varepsilon^c_{k,\Lambda}&= \bar \varepsilon^c_{k}+U n^f  , &
\Delta^{}_{k,\Lambda} &= 0^+ ,\nonumber \\
\varepsilon^f_{k,\Lambda}&= \bar\varepsilon^f_{k}+U n^c  , &
U^{}_{k_1k_2k_3,\Lambda} &= U.
\label{5.16}
\end{align}
Note that the energy shift $E_\lambda$ in ${{\mathcal H}_{0, \lambda}}$ has 
no effect on expectation values and will again be left out in what follows.

Next we have to construct the generator $X_{\lambda, \Delta \lambda}$ of 
transformation \eqref{2.8}. In the minimal transformation and in lowest order perturbation theory according to Eqs.~\eqref{2.13} and \eqref{5.15} the generator
would read,
\begin{equation}
\label{5.17}
\begin{split}
{\mathbf Q}_{\lambda - \Delta \lambda} {X}_{\lambda, \Delta \lambda} &=
\sum_{{k}_1{k}_2{k}_3}
\frac{U^{}_{k_1k_2k_3,\lambda}}
 {{\omega}^{}_{{k}_1{k}_2{k}_3,\lambda}} \, 
\big(1- \Theta_{{k}_1{k}_2{k}_3, \lambda- \Delta \lambda }
\big) \\ 
&\hspace*{1.0cm}\times\Theta_{{k}_1{k}_2{k}_3, \lambda}
:a^{}_{k_1k_2k_3}:,
\end{split}
\end{equation}
where we have defined
$
\Theta_{{k}_1{k}_2{k}_3, \lambda} =
\Theta(\lambda-|{\omega}^{}_{{k}_1{k}_2{k}_3,\lambda}|)
$. In Eq.~\eqref{5.17} the product of the two $\Theta$-functions assures that only excitations between 
$\lambda - \Delta \lambda$ and $\lambda$ are eliminated by the unitary transformation \eqref{2.8}.

Instead, 
we use as generator the part
\begin{equation}
\label{5.18}
\begin{split}
{\bf P}_{\lambda- \Delta \lambda}{X}_{\lambda,\Delta\lambda}&= 
\frac{1}{N}\sum_{k_1k_2k_3}
{A}^{}_{k_1k_2k_3,\lambda} \,
\Theta_{{k}_1{k}_2{k}_3, \lambda- \Delta \lambda}\\ 
&\hspace*{0.5cm} \times
\Theta_{{k}_1{k}_2{k}_3, \lambda} \,
:a^{}_{k_1k_2k_3}:,
\end{split}
\end{equation}
where the coefficients $A^{}_{k_1k_2k_3,\lambda}$ are chosen proportional to 
$\Delta \lambda$, $A^{}_{k_1k_2k_3,\lambda} =\Delta \lambda \
\alpha^{}_{k_1k_2k_3,  \lambda}$,
with 
\begin{equation}
\label{5.20}
\alpha^{}_{k_1k_2k_3, \lambda} 
=\frac{\omega^{}_{{k}_1{k}_2{k}_3,\lambda}}
{\kappa(\lambda-|{\omega}_{{k}_1{k}_2{k}_3,\lambda}|)^2}\, U^{}_{k_1k_2k_3,\lambda}.
\end{equation}
As shown in Subsec.~\ref{App A1b} expression \eqref{5.18} with \eqref{5.20} is an appropriate choice in the continuous version of our generalized diagonalization scheme, where the operator structure is taken over from Eq.~\eqref{5.17}.
The two $\Theta$-functions guarantee that Eq.~\eqref{5.18} is  
the generator part with low energy excitations only, $|{\omega}_{{k}_1{k}_2{k}_3,\lambda}|< \lambda$ 
and $|{\omega}_{{k}_1{k}_2{k}_3,\lambda- \Delta \lambda}|< \lambda -\Delta \lambda$.  
The constant $\kappa$ in (\ref{5.20}) again denotes an energy constant to ensure that the 
coefficients $A^{}_{k_1k_2k_3, \lambda}$ 
are dimensionless.

The next step is to derive renormalization equations for the Hamiltonian $\mathcal H_\lambda$. They are obtained from the perturbative expression \eqref{2.17} for $\mathcal H_{\lambda- \Delta \lambda}$ by identifying $X^{(1)}_{\lambda, \Delta \lambda}$ with  \eqref{5.18}
(and setting $X^{(2)}_{\lambda, \Delta \lambda}$ equal to zero). Then Eq.~\eqref{2.17} reduces to 
\begin{equation*}
\begin{split}
\mathcal H_{\lambda -\Delta\lambda} &= \mathcal H_\lambda + {\mathbf P}_{\lambda- \Delta \lambda} [X_{\lambda,\Delta \lambda},  \mathcal H_{0,\lambda} +  \mathcal H_{1,\lambda} \big]  \\
&+ \frac{1}{2} {\mathbf P}_{\lambda- \Delta \lambda} \big[ \big[ X_{\lambda,\Delta \lambda}, [ X_{\lambda,\Delta \lambda},  \mathcal H_{0,\lambda}  \big] \big] .
 \end{split}
  \end{equation*}
Since the last term is of second order in $\Delta \lambda$, it vanishes in the limit $\Delta \lambda \rightarrow 0$. Then,  the derivative of $\mathcal H_\lambda$ with respect to $\lambda$ becomes 
\begin{equation*}
\frac{d{\mathcal H}_\lambda}{d\lambda} =-
\frac{1}{N}\sum_{k_1k_2k_3}
\alpha^{}_{k_1k_2k_3,\lambda} 
\Theta_{{k}_1{k}_2{k}_3, \lambda} 
[ :a^{}_{k_1k_2k_3}: \, ,  {\mathcal H}_{\lambda} ],
\end{equation*}
where Eqs.~\eqref{5.18} and \eqref{5.20} have been used. 
To find the renormalization equations for the $\lambda$-dependent parameters of $\mathcal H_\lambda$  
the commutator on the right hand side has to be evaluated. As for the Holstein model, one is also led to new operator expressions which are 
 not present in {\em ansatz}~\eqref{5.11}, \eqref{5.12} for $\mathcal H_\lambda$. 
 Therefore, again  an additional factorization of the form of Eqs.~\eqref{4.5} has to be applied in order to trace back all operator structures to those 
 present in $\mathcal H_\lambda$.  Finally, comparing the result with the generic 
 expression of $\mathcal H_\lambda$, given by  Eqs.~\eqref{5.11}, \eqref{5.12}, 
 one finds the desired set of coupled renormalization equations. For example the renormalization equation for $\varepsilon^c_{k,\lambda}$ reads:
  \begin{align}
 \label{5.22}
\frac{d\varepsilon^c_{k,\lambda}}{d\lambda}=
-&\frac{1}{N^2}\sum_{k_1k_2}U^{}_{k_1kk_2,\lambda}
\alpha^{}_{kk_1,k_1+k_2-k,\lambda}
(1- n^c_{k_1} )\nonumber\\
&\times( n^f_{k_1+k_2-k}-  n^f_{k_2} )\nonumber\\
-&\frac{1}{N^2}\sum_{k_1k_2}U^{}_{kk_1k_2,\lambda}
\alpha^{}_{k_1k,k+k_2-k_1,\lambda}
n^c_{k_1} \nonumber\\
&\times( n^f_{k+k_2-k_1} -  n^f_{k_2}),
\end{align}
where we have defined expectation values $n_{k}^c =  \langle c_{k}^\dagger  c_{k}\rangle$ and $n_{k}^f =  \langle f_{k}^\dagger  f_{k}\rangle$ which are formed with the full Hamiltonian $\mathcal H$. Similar equations are found for the remaining parameters $\varepsilon^f_{k,\lambda}$ and $\Delta_{k,\lambda}$. The additional renormalization equation for the 
$\lambda$-dependence of the coupling $U^{}_{k_1k_2k,\lambda}$  reads 
\begin{eqnarray}
\label{5.26}
\frac{d U^{}_{k_1k_2k,\lambda}}{d\lambda} =
{\omega}^{}_{k_1k_2k,\lambda} 
\alpha^{}_{k_1k_2k,\lambda}.
\end{eqnarray}
Having the structure \eqref{5.20} of $\alpha_{k_1k_2k_3,\lambda}$ in mind 
one might expect that $U_{ k_1  k_2  k_3, \lambda}$ diverges at $\lambda= \omega_{ k_1 k_2 k_3, \lambda}$.
However, as it turns out, it vanishes exponentially at this point which follows from the renormalization equation \eqref{5.26} for  
$U_{ k_1 k_2 k_3, \lambda}$ together with Eq.~\eqref{5.20}  (also compare Subsec.~\ref{A.1}). Thus, we arrive at a free model.
Integrating the whole set of differential equations with the initial values given by 
Eqs.~\eqref{5.16}, the completely renormalized Hamiltonian $\tilde{\mathcal H}:=
{\mathcal H}_{\lambda \rightarrow 0}={\mathcal H}_{0, \lambda \rightarrow 0}$
becomes 
\begin{equation}
\begin{split}
\label{5.27}
\tilde{\mathcal{H}}&=
 \sum_{k}\tilde{\varepsilon}^c_{k}:c^\dagger_{k}c^{}_{k}:
+\sum_{k}\tilde{\varepsilon}^f_{k}:f^\dagger_{k}f^{}_{k}: \\
&+ \sum_{k}(\tilde{\Delta}^{}_{k}:f^\dagger_{k}c^{}_{k}:+\,\mbox{h.c.}) .
\end{split}
\end{equation}
Again the quantities with tilde sign denote the parameter values at $\lambda = 0$. 
The final Hamiltonian \eqref{5.27} can be diagonalized by use of the transformation \eqref{B21} leading to 
\begin{equation}
\label{5.28}
{\tilde{\mathcal H}}
=\sum_{{k}}E^c_{{k}}:\bar{c}^\dagger_{{k}}\bar{c}^{}_{{k}}:
+\sum_{{k}}E^f_{{k}}:\bar{f}^\dagger_{{k}}\bar{f}^{}_{{k}}:
+\tilde{E}. 
\end{equation}
Here $\bar c_k^\dag = u_k c^\dag_k + v_k f^\dag_k$ and $\bar f_k^\dag = -v_k c^\dag_k + u_k f^\dag_k$ 
are the quasi-particle operators which are linear combinations of the old $c$- and $f$-operators.
As in Subsec.~\ref{III.A} all expectation values appearing in the set of renormalization equations have to be evaluated 
self-consistently.  According to relation \eqref{2.21}, the same unitary transformation as 
for the Hamiltonian has to be applied to the operators. For instance, 
following Eq.~\eqref{2.21}, the expectation value $ n^c_{k}$ can be expressed by $ n^c_{k} =\langle c^\dagger_{k, \lambda= 0} 
c^{}_{k, \lambda = 0} \rangle_{\tilde{\mathcal{H}}}$
where the average on the right hand side  is now formed with the fully renormalized Hamiltonian $\tilde{\mathcal{H}}$, 
 and $c_{k, \lambda}$ is given by
$c_{k, \lambda} =  e^{{X}_\lambda} c_{k}
e^{- {X}_\lambda}$. For the transformed operator we use the following {\em ansatz},
\begin{align*}
:c^\dagger_{k, \lambda}&:= x^{}_{k,\lambda}:c^\dagger_{k}: \\
+&\frac{1}{N^2}\sum_{k_1k_2}
y^{}_{k_1kk_2,\lambda}:c^\dagger_{k_1}
f^\dagger_{k_2}f^{}_{k_1+k_2-k}:. 
\end{align*}
Here, the operator structure is again taken over from the lowest order expansion in $X_{\lambda,\Delta \lambda}$
of the unitary transformation \eqref{2.7}. For the $\lambda$-dependent coefficients 
$x^{}_{k,\lambda}$ and $y^{}_{k_1kk_2,\lambda}$ new
 renormalization equations have to be derived. In analogy to the former derivation 
 for $\mathcal H_\lambda$ one finds for example
 \begin{equation}
\frac{dy^{}_{k_1kk_2,\lambda}}{d\lambda}=-x^{}_{k,\lambda}
{\alpha}^{}_{k_1kk_2,\lambda}.
\label{5.36}
\end{equation}
and a similar differential equation for $x^{}_{k,\lambda}$.
Integration between $\Lambda$ (where $x^{}_{k,\Lambda}=1$ 
and $y^{}_{k_1kk_2,\Lambda}=0$) and $\lambda =0$ leads to 
\begin{align}
\label{5.37}
:c^\dagger_{k, \lambda = 0}&:=\tilde{x}^{}_{k}:c^\dagger_{k}:\nonumber\\
+&\frac{1}{N^2}\sum_{k_1k_2}
\tilde{y}^{}_{k_1kk_2}:c^\dagger_{k_1}
f^\dagger_{k_2}f^{}_{k_1+k_2-k}:,
\end{align}
from which $ n_{k}^c $
is found,
\begin{equation}
\begin{split}
\label{5.38}
 n^c_{k} &=
|\tilde{x}^{}_{k}|^2\langle c^\dagger_{k}c^{}_{k}\rangle_{\tilde{\mathcal{H}}}\\
&+\frac{1}{N^2}\sum_{k_1k_2}\left|
\tilde{y}^{}_{k_1kk_2}\right|^2
\langle c^\dagger_{k_1}c^{}_{k_1}\rangle_{\tilde{\mathcal{H}}}
\langle f^\dagger_{k_2}f^{}_{k_2}\rangle_{\tilde{\mathcal{H}}}\\
&\times\left(1-\langle f^\dagger_{k_1+k_2-k}
f^{}_{k_1+k_2-k}\rangle_{\tilde{\mathcal{H}}}\right).
\end{split}
\end{equation}

 The remaining  expectation values from Eq.~\eqref{5.38} and from an analogous equation for $n^f_k$
 are found from a similar {\em ansatz} for 
$:f^\dagger_k(\lambda):$.
In the last step the expectation values
on the right hand  side of Eq.~\eqref{5.38} are needed. 
With the diagonal form of $\tilde{\mathcal H}$ in Eq.~\eqref{5.28}
one finds $\langle c^\dagger_{k}c^{}_{k}\rangle_{\tilde{\mathcal H}}
=u^2_{k}f(E^c_{k})+v^2_{k}f(E^f_{k})$ and a corresponding expression for the $f$-electrons.
Here $f(E^{c(f)}_{k})$ are Fermi functions and the coefficients $u_k$ and $v_k$ are related via Eqs.~\eqref{A.5} to the original dispersions.

As a second example for expectation values, let us consider the one-particle spectral function  for $c$-electrons, $A^{c}({k},\omega)=(1/\pi)\textrm{Im}G^c({k},\omega)$,
where $G^c({k},\omega) = i  \int_{0}^\infty  dt \, 
\langle  [c^{}_{k}(t) , c^{\dagger}_k ]_+ \rangle^{} \, e^{i (\omega + i\eta) t}$ is the Fourier transform of the retarded Green's function 
($\eta =0^+$). Note that, in contrast to the occupation numbers considered so far, the spectral function involves also dynamical properties of the system. 
Using again relation \eqref{2.21}, the spectral function can be rewritten as $G^c(k,\omega)= i  \int_{0}^\infty  dt \, 
\langle  [c^{}_{k, \lambda =0}(t) , c^{\dagger}_{k, \lambda=0} ]_+ \rangle^{}_{\tilde{\mathcal H}} \,
e^{i (\omega + i \eta) t}$,
where the expectation value on the right hand side and the time dependence are again formed with 
$\tilde{\mathcal H}$.  
With  expression \eqref{5.37} for $c^\dagger_{k, \lambda = 0}$, 
we are immediately led to the following result for the 
$c$-electron spectral function, 
\begin{align}
\label{5.48}
A^{c}&(k,\omega)=|\tilde{x}^{}_{k}|^2\left[u^2_{k}\delta(\omega-E^c_{k})+v^2_{k}\delta(\omega-E^f_{k})\right]\nonumber\\
+&\frac{1}{N^2}\sum_{k_1k_2}|\tilde{y}^{}_{k_1kk_2}|^2\delta\left(\omega-(E^c_{k_1}-E^f_{k_1+k_2-k}+E^f_{k_2})\right)\nonumber\\
&\times\Big[\langle c^\dagger_{k_1}c^{}_{k_1}\rangle_{\tilde{\mathcal{H}}}\left(\langle f^\dagger_{k_2}f^{}_{k_2}\rangle_{\tilde{\mathcal{H}}}-
\langle f^\dagger_{k_1+k_2-k}f^{}_{k_1+k_2-k}
\rangle_{\tilde{\mathcal{H}}}\right)\nonumber\\
&+\langle f^\dagger_{k_1+k_2-k}f^{}_{k_1+k_2-k}
\rangle_{\tilde{\mathcal{H}}}\left(1-\langle f^\dagger_{k_2}f^{}_{k_2}\rangle_{\tilde{\mathcal{H}}}\right)\Big].
\end{align}
The  structure of the poles in the  first term of  $A^{c}(k,\omega)$ and also  in $A^{f}(k,\omega)$ describes coherent excitations whereas the second term gives rise to incoherent contributions. In the same way one can also calculate the spectral function $A^{f}(k,\omega)$ for $f$-electrons.   


The analytical expressions for $A^{c}(k,\omega)$ and also for $A^{f}(k,\omega)$
outlined so far were evaluated numerically in Ref.~\cite{PBF2010}. Here we shortly review the technical  
procedure of  this analysis and discuss in the following only one representative result. 
Thereby, the main task is the numerical solution of the set of  the coupled renormalization
equations. 
To this end,  some initial values for $n^c_{k}, \dots$ are chosen and then 
 the renormalization of the Hamiltonian and of all other operators is determined
by solving  the differential 
equations~\eqref{5.22}, \eqref{5.26}, \eqref{5.36}, and the equations for the remaining parameters. Integrating these equations between $\lambda =\Lambda$ and $\lambda =0$, 
all model parameters will be renormalized. 
Finally,  using $\tilde{\mathcal{H}}$, the new expectation 
values given for example by Eq.~\eqref{5.38}  are calculated  and the renormalization 
process of the Hamiltonian is restarted.

\begin{figure}[t]
\centering
\includegraphics[width=1\columnwidth] {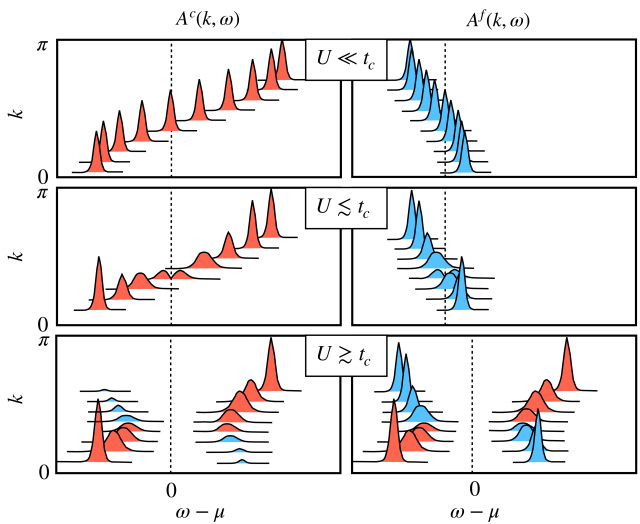}
\caption{(Color online) Schematic picture of the zero temperature results for the $1d$ spectral functions 
$A^c(k, \omega)$ (left panels) and  $A^f(k, \omega)$ (right panels) at half-filling for several  characteristic values of $U$. Red and blue colors indicate whether the excitations are of '$c$' or '$f$' character. The actual numerical results can be found in Ref.~\cite{PBF2010}. 
For small $U \ll t_c$ the spectral functions are dominated by the 'coherent' contributions caused by the first term in Eq.~\eqref{5.48}  and  in the equation for $A^f(k,\omega)$. 'Incoherent' contributions are negligible. For increasing values of $U \approx t_c$  a gap begins to open at the Fermi momentum caused by incoherent contributions. For even larger $U \gg t_c$  the gap broadens further. Of particular importance is a considerable admixture of $c$ electron contributions to the $f$ electron spectrum between $k=0$ and $k\approx k_F$. Note that the vertical dotted lines mark the chemical potential $\mu$. Further note that at medium values of $U \gtrsim t_c$ a gap opens at $\mu$. }
\label{fig:Awcfef10}
\end{figure}
%

In Fig.~\ref{fig:Awcfef10} a qualitative picture of the zero-temperature, wave-vector, and energy resolved single-particle spectral functions $A^c (k, \omega)$ (left panels)
and $A^f(k, \omega)$ (right panels) is shown for several characteristic values of $U$ for half-filling according to Ref.~\cite{PBF2010}.  The two different colors indicate the respective character of the excitations (red: $c$-like, blue: $f$-like).
The bare band structure has been taken over from Ref.~\cite{PBF2010} where $\varepsilon^f=-1$
($\varepsilon^c=0$), $t^f=-0.3$ ($t^c=1$).
For weak Coulomb interaction (upper panels) 
the system is in a semi-metallic phase, and consistently
$A^c(k,\omega)$ and $A^f(k,\omega)$ follow the nearly unrenormalized
$c$- and $f$-band dispersions, respectively. Concomitantly, 
 a more or less uniform distribution of the spectral 
weight is found and incoherent contributions to
$A^c(k,\omega)$ and $A^f(k,\omega)$ can be neglected. By increasing $U$ to some value near $t_c$ (intermediate regime) a new phase is entered where a gap feature develops at the
Fermi energy (Fermi momentum), but away from that the spectra 
still show the main characteristics of the semi-metallic state 
(cf. both middle panels). At very large $U \gg t_c$, the gap broadens.
Most notably, however, is a small redistribution 
of spectral weight from the coherent to the incoherent part, which is mostly seen in  $A^f(k,\omega)$, with pronounced absorption maxima
at $k=0,\,\pi$. It leads to a considerable admixture of 
$c$-electron contributions to the $f$-electron spectrum. Note that this property does not follow from a possible transfer of spectral weight from $u_k^2$ to $v_k^2$ within the coherent part of $A^c(k,\omega)$ and $A^f(k,\omega)$ (first line of Eq.~\eqref{5.48}) as it is shown numerically in Ref.~\cite{PBF2010}. The resulting double peak structure around the 
Fermi level can be considered as an almost $k$-independent bound object of 
$c$ electrons and $f$ holes.  




\section{Conclusion}
\label{X}

We have presented a generalized diagonalization scheme  
for many-particle systems which includes both the previously introduced 
projector-based renormalization method and also Wegner's flow equation method as special cases. Instead of 
eliminating high-energy states as in usual renormalization group 
methods in the present approach high-energy transitions
are successively eliminated using a sequence of unitary
transformations. Thereby,  all states of the 
unitary space of the interacting system are kept. 
In that respect, the presented method is closely
related to the known similarity transformation introduced by G\l azek and Wilson.
   
The method starts from a Hamiltonian
which is decomposed into a solvable unperturbed part 
and a perturbation, ${\cal H} = {\cal H}_0 +
{\cal H}_1$, where the latter part is responsible for transitions 
between the eigenstates of ${\cal H}_0$. 
Suppose a renormalized Hamiltonian ${\cal H}_\lambda$ has been constructed
in such a way that all transitions with  
energies larger than some cutoff energy $\lambda$ are already eliminated.  
Then, ${\cal H}_\lambda$  
can further be renormalized by eliminating also all transitions from the energy shell 
between $\lambda$ and a somewhat reduced 
cutoff $\lambda - \Delta \lambda$ leading to a renormalized Hamiltonian $\mathcal H_{\lambda - \Delta \lambda}$.  Based on   
a unitary transformation, $\mathcal H_{\lambda - \Delta \lambda} =
e^{X_{\lambda,\Delta \lambda}}\, {\cal H}_\lambda\, 
e^{-X_{\lambda,\Delta \lambda}}$, it is guaranteed  that both $\mathcal H_\lambda$ and 
$\mathcal H_{\lambda - \Delta \lambda}$  have the same eigenspectrum. 
The generator 
$X_{\lambda, \Delta \lambda}$ is specified by the condition 
${\bf Q}_{\lambda - \Delta \lambda}{\cal H}_{\lambda - \Delta \lambda}=0$
where  ${\bf Q}_{\lambda - \Delta \lambda}$ is the projector on all 
transitions with energies larger than 
$\lambda - \Delta \lambda$. The latter condition implies that 
all transitions from the shell between $\lambda$ and $\lambda - 
\Delta \lambda$ are eliminated which leads to the renormalization of  
${\cal H}_{\lambda - \Delta \lambda}$. Thereby only 
the part ${\bf Q}_{\lambda - \Delta \lambda} X_{\lambda, \Delta \lambda}$ of the generator is fixed whereas 
the orthogonal part  ${\bf P}_{\lambda - \Delta \lambda}
X_{\lambda, \Delta \lambda}$ can still be chosen arbitrarily.  By proceeding the renormalization up 
to the final cutoff $\lambda =0$ all transitions from ${\cal H}_{1,\lambda}$
are successively eliminated. Correspondingly, the fully renormalized Hamiltonian 
$\tilde{\cal H}={\cal H}_{0,\lambda=0}$ is diagonal and allows in principle
to evaluate any correlation function of physical interest. This property even allows to consider realistic parameters of a particular material and to compare the numerical results of the renormalization equations with the experiment on a quantitative level. 
In particular, the one-particle excitations of $\tilde{\cal H}$ can be 
considered as quasi-particles of the coupled many-particle system since the 
eigenspectrum of the original interacting Hamiltonian ${\cal H}$ 
and of  $\tilde{\cal H}$ are the same as both are connected by 
unitary transformations.  Finally, one should add that  one fundamental advantage of the presented approach is that it allows to interpret all features of the renormalization on the basis of the renormalization equations.

The additional freedom in the choice of the  remaining part ${\bf P}_{\lambda - \Delta \lambda}
X_{\lambda, \Delta \lambda}$ of the generator $X_{\lambda, \Delta \lambda}$ can be used in a different way.
Whereas in the PRM this part was mostly set equal to zero ('minimal' choice), there is a connection to Wegner's 
continuous flow equation method. In this method  ${\bf P}_{\lambda - \Delta \lambda}
X_{\lambda, \Delta \lambda}$ is chosen such that the relevant interaction parameters decay exponentially
in the flow with decreasing $\lambda$. In this way, when the interactions have vanished, one also ends up      
with a free model that can be solved. Note that in Wegner's method the renormalization depends on  an 
appropriate choice of ${\bf P}_{\lambda - \Delta \lambda} X_{\lambda, \Delta \lambda}$. In contrast, in the 'minimal' transformation
the generator  ${\bf Q}_{\lambda - \Delta \lambda} X_{\lambda, \Delta \lambda}$ is uniquely fixed in the 
formalism and does not rely on a reasonable choice of the generator. Moreover,  
the stepwise transformation of the PRM allows
to describe the physical behavior
on both sides of a quantum critical point, taking symmetry breaking terms in the 'unperturbed' part
${\cal H}_{0,\lambda}$ into account.  Thereby the transformation of 
eigenmodes of the Liouville operator 
${\bf L}_{0,\lambda}$ can be followed in each renormalization step, which 
makes the study of quantum critical points possible. 
  
In the numerical evaluation of the renormalization step between $\lambda$ and $\lambda -\Delta \lambda$ 
the width $\Delta \lambda$ should be  chosen sufficiently small so that
 only a small  number of renormalization processes contribute
  within the interval $\Delta \lambda$. 
 Therefore, the  'smallness parameter' for the expansion of the transformation  $\mathcal H_{\lambda - \Delta \lambda} =
e^{X_{\lambda,\Delta \lambda}}\, {\cal H}_\lambda\, 
e^{-X_{\lambda,\Delta \lambda}}$  is  given by
 the relative coupling parameter of $\mathcal H_1$ multiplied by the  small 
 ratio of the number of  renormalization processes within $\Delta \lambda$ divided by the total number of all processes. 
 In this way, perturbation theory in $X_{\lambda, \Delta \lambda}$ for the $\Delta \lambda$ steps should be well fulfilled leading to 
reliable  results for coupling parameters up to  the order of the scale of $\mathcal H_0$.
  
Finally, let us find out what  possibilities are there  to further develop the diagonalization scheme in order to access additional fields of application. Problems for the method may arise if the eigenvalue problem of ${\cal H}_{0,\lambda}$ can not be solved exactly and additional approximations become necessary which might cause uncontrolled errors in the renormalization processes. One possibility to circumvent such problems would be to use at first an alternative many-particle approach which maps the original Hamiltonian to an effective model with a solvable unperturbed part which, however, still contains interactions. A subsequent application of the approach to the effective model could  integrate out the remaining interactions. In this way an extremely powerful tool is found which also could treat strongly correlated systems. Thereby, after having identified the dominant renormalization processes a direct access to the experimental observables is achieved. 

A prominent example for such a methodical combination could be the strongly correlated Hubbard model, though an exclusive application of the diagonalization technique to eliminate the Hubbard interaction should also be possible.  However, in this case one is led to an increasing number of complicated local interaction operators in ${\cal H}_{0,\lambda}$ preventing the solution of its eigenvalue problem. One possible way to circumvent this difficulty is to combine our method with the dynamical mean-field theory (DMFT). The DMFT tackles the many-particle problem by using a self-consistency cycle with the aim of obtaining an effective impurity Green’s function which corresponds to the many-particle Green’s function in the limit of infinite dimensions \cite{DMFT_2}. However, an appropriate method to solve the corresponding impurity problem must be implemented in the DMFT process. The usual way of handling this step is to solve an effective Anderson impurity model (AIM) consisting of  an auxiliary bath of conduction electrons which hybridizes with the impurity states. However, its solution is a highly non-trivial task and requires sophisticated numerical treatment. An alternative and more efficient approach to the AIM could be provided by our proposed diagonalization scheme since it has  significant advances over other methods due to its analytical nature. Our method could be applied to integrate out the hybridization coupling between bath and impurity leading to an effectively free system of the bath fermions and the impurity. This would allow us not only to provide the impurity Green’s function for the  DMFT loop but could be at the same time  used to directly calculate physical quantities of interest, as for example transport coefficients, spectral functions, or susceptibilities. Moreover, compared to purely numerical methods our analytical theory can handle much larger system sizes.

On the other hand, for low-dimensional systems the DMFT step will introduce additional approximations, since the usual DMFT is exactly valid only in infinite dimensions. However, the proposed combination of the DMFT with our diagonalization method could reduce the impact of this approximation in the following way. Compared to conventional methods to solve the impurity problem our approach would naturally allow us to take into account non-local extensions of the DMFT. Such extensions could be for example the replacement of the single-site impurity by a cluster, which includes non-local correlations within the cluster\cite{Hettler1998,Kotliar2001}. It results in a better description of low-dimensional systems or systems close to the Mott transition\cite{Toschi2007}. Our proposed method might be able to  handle this correction by simply taking into account a momentum-dependence in the hybridization function and the bath system, which would not cause any additional difficulties. Therefore, we believe that the effect of approximations from the intermediate DMFT step could be reduced by a proper extension of the effective AIM. In conclusion, combining the two methods should provide an extremely powerful tool to treat strongly correlated materials with on-site interactions as in the Hubbard model including an interpretation of relevant 
experiments.

A second example for a possible combination with other methods  is provided by material-specific ab initio methods as quantum chemistry calculations or results from the density functional theory which would lead to  the initial parameters for a subsequent application of the diagonalization method. Thus, combined with the numerical methods our approach may provide band structure, Hubbard and exchange interactions of real materials.

\section*{Acknowledgements}
We would like to thank J.~van den Brink, H.~Fehske,
J.~Geck, C.~Hess, V.-N.~Phan, B.~Buchholz, and J.~Trinckauf for helpful discussions.

\paragraph*{Funding information}
This project has received funding from the European Research Council (ERC) under the European Unions Horizon 2020 research and innovation programme (grant agreement No 647276 – MARS – ERC-2014-CoG). S.S.~acknowledges financial support by the Deutsche Forschungsgemeinschaft via the Emmy Noether Programme ME4844/1-1 (project id 327807255), the Collaborative Research Center SFB 1143 (project id 247310070), and the Cluster of Excellence on Complexity and Topology in Quantum Matter \textit{ct.qmat} (EXC 2147, project id 390858490).


\end{document}